# Which Ion Dominates Temperature and Pressure Response of Halide Perovskites and Elpasolites?

Loreta A. Muscarella,[1#*] Huygen J. Jöbsis,[1#] Bettina Baumgartner,[1] P. Tim Prins,[1] D. Nicolette Maaskant,[1] Andrei V. Petukhov,[2] Dmitry Chernyshov,[3] Charles J. McMonagle,[3] and Eline M. Hutter[1*]

1. Inorganic Chemistry and Catalysis group, Debye Institute for Nanomaterials Science and Institute for Sustainable and Circular Chemistry, Department of Chemistry, Utrecht University, Princetonlaan 8, 3584 CB Utrecht, the Netherlands
2. Physical and Colloid Chemistry, Debye Institute for Nanomaterials Science, Department of Chemistry, Utrecht University, Padualaan 8, 3584 CH Utrecht, the Netherlands
3. Swiss–Norwegian Beamlines, European Synchrotron Radiation Facility, 71 Avenue des Martyrs, 38000 Grenoble, France

# These authors contributed equally

* Correspondence should be addressed to E.M.Hutter@uu.nl and loretaangela.muscarella@gmail.com

**Abstract**

Halide perovskite and elpasolite semiconductors are extensively studied for optoelectronic applications due to their excellent performance together with significant chemical and structural flexibility.

However, there is still limited understanding of their basic elastic properties and how they vary with composition and temperature, which is relevant for synthesis and device operation. To address this, we performed temperature- and pressure-dependent synchrotron-based powder X-ray diffraction (XRD). In contrast to previous pressure-dependent XRD studies, our relatively low pressures (ambient to 0.06 GPa) enabled us to investigate the elastic properties of halide perovskites and elpasolites in their ambient crystal structure. We find that halide perovskites and elpasolites show common trends in the bulk modulus and thermal expansivity. Both materials become softer as the halide ionic radius increases from Cl to Br to I, exhibiting higher compressibility and larger thermal expansivity. The mixed-halide compositions show intermediate properties to the pure compounds. Contrary, cations show a minor effect on the elastic properties. Finally, we observe that thermal phase transitions in *e.g.,* $MAPbI_3$ and $CsPbCl_3$ lead to a softening of the lattice, together with negative expansivity for certain crystal axes, already tens of degrees away from the transition temperature. Hence, the range in which the phase transition affects thermal and elastic properties is substantially broader than previously thought. These findings highlight the importance of considering the temperature-dependent elastic properties of these materials, since stress induced during manufacturing or temperature sweeps can significantly impact the stability and performance of the corresponding devices.

**Table of Contents**

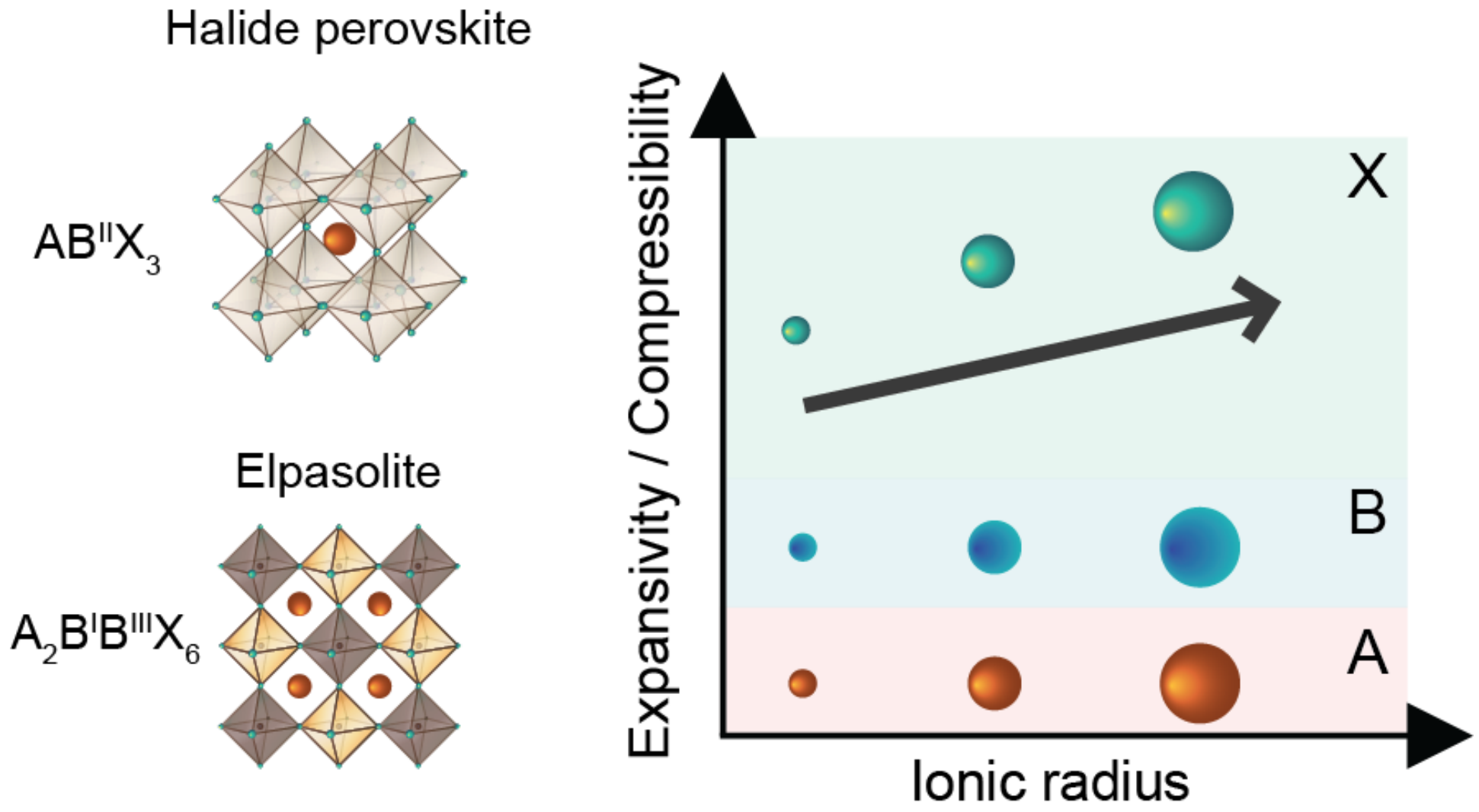

## Introduction

Lead halide perovskite and related lead-free elpasolite (*i.e.,* double perovskite) semiconductors are widely investigated for a variety of optoelectronic applications, including photovoltaics,[1–4] light emitting diodes,[5,6] photoredox catalysis,[7–9] and radiation detection.[10–12] The unprecedented interest in this class of materials can be understood from its excellent optoelectronic performance, in combination with enormous chemical and structural flexibility. For instance, mixing halides in lead halide perovskites enables tuning the absorption onset and emission wavelength over the entire visible range of the light spectrum.[13] In halide elpasolites, with the general formula $A_2B^IB^{III}X_6$, bandgap tunability can be obtained by mixing metals at the $B^{III}$-site, such as Bi, Sb, In or Fe,[14–16] and to some extent by mixing halides (X-site).[17–19] In contrast to the plethora of studies on optoelectronic properties and applications of halide perovskites and elpasolites, there is still limited knowledge about their mechanical properties, such as compressibility and expansivity, and how these properties vary with temperature. Additionally, little is known about the temperature-dependent crystal phases of these materials. The solution-processed deposition of halide perovskites and elpasolites in the form of thin films onto substrates involves elevated temperatures ranging from 100 to 200°C. Additionally, the differences in the structural and thermal properties of substrates and halide perovskites, such as lattice parameters, thermal expansion coefficient, can lead to strain and deformations during the thin film deposition. Previously, we and others have found that compression slows down halide migration, and that compression or strain can activate or suppress light-induced halide segregation.[20–23] Thus, quantifying to what extent this class of relatively soft materials responds to changes in temperature and pressure leading to stress-induced deformations, is key for growing stable thin films of these perovskites and manipulating their optoelectronic properties. Previous studies that report on elastic properties used diamond anvil cells (DACs) to apply external pressure,[22,24–30] and hence probed high pressure regimes (several to hundreds GPa) and crystal phases that are far from relevant to their ambient crystal structure. The deformations induced in halide perovskite films during the solution-processed deposition are comparable to the exertion of mild pressure (<0.5 GPa).[31] Therefore, hydraulic pressure techniques are needed to investigate how the crystal structure would respond in the more relevant pressure range. In this synchrotron-based powder X-ray diffraction (XRD) study, we investigated the structural properties of halide perovskites and elpasolites at pressures between 0.004 and 0.060 GPa, using a hydrostatic pressure cell. Importantly, this allowed us to determine the elastic properties of the materials in their ambient crystal structures. The pressure-dependent measurements were performed between room temperature and 90°C (298 to 363 K), a typical temperature range for thin film synthesis and photovoltaic operation conditions. We report the temperature-dependent bulk modulus for a wide variety of compositions, including $MAPbCl_3$, $-Br_3$, and $-I_3$ (MA = methylammonium), mixed halide variants thereof, $CsPbCl_3$, $-Br_3$, and $-I_3$, and the elpasolites $Cs_2AgBiCl_6$, $-Br_6$, mixed halide and trivalent metal variants thereof. Hence, we find some general trends in elastic properties of halide perovskites and elpasolites. For all perovskite and elpasolite materials, we find that the iodide-based materials are substantially softer than their bromide- and chloride-based analogues, and that the bulk modulus increases from $I^-$ to $Br^-$ to $Cl^-$. The mixed-halide perovskites show bulk moduli in between their pure compounds and, within the same crystal phase, exhibit a linear relation with the average halide radius. In addition, the bulk modulus appears constant in the investigated temperature range for all materials, provided that the materials remain in the same crystal structure. For both $CsPbCl_3$ and $MAPbI_3$, that undergo phase transitions at around 325 and 330 K respectively, we observe a reduction of the bulk modulus at temperatures close to the phase transition. In addition, as the elastic properties of non-cubic perovskites area anisotropic, we also estimate the compressibility in different crystallographic directions. Finally, we find that some temperature-dependent phase transitions of halide perovskites are reflected in negative thermal expansion coefficients of certain crystal axes, already several tens of degree below the transition temperature. The variation of thermal expansivity with crystal axis (for non-cubic perovskites) and crystal structure (*i.e.,* temperature) could induce stress during synthesis or temperature cycling, and in turn affect the stability of the corresponding devices (*e.g.*, ion migration). Therefore, elastic properties should be considered for thin films when designing optoelectronic devices of halide perovskites.

## Results and Discussion

Microcrystalline powder samples of $MAPb(Cl_{1-x}Br_x)_3$, $MAPb(I_{1-x}Br_x)_3$, $CsPbCl_3$, $-Br_3$, and $-I_3$, and several elpasolites (based on $Cl^-$, $Br^-$, $Bi^{3+}$ and/or $Fe^{3+}$, $Sb^{3+}$, $In^{3+}$) were made via mechanochemical synthesis in a ball mill, as described in the **Experimental Section** in the **Supporting Information**. To quantify the effect of both pressure and temperature on the structural parameters of these halide perovskites and elpasolites, we performed pressure- and temperature-dependent powder XRD measurements using a synchrotron radiation source at the European Synchrotron Radiation Facility (ESRF) beamline BM01.[32] We used a hydrostatic pressure cell filled with a fluorinated inert liquid (FC-770) as a pressure-transmitting medium and we performed pressure sweeps from 0.004 to 0.060 GPa at temperatures of 298, 318, 335, and 355 K. Additional details on the experimental setup can be found in the **Experimental Section** of the **Supporting Information**. This approach enables to directly measure the variation of structural parameters as a function of pressure and temperature and thereby allows to determine material properties such as the bulk modulus (B), the compressibility ($K$), and thermal expansivity ($\alpha$). **Figure 1a** shows selected reflections of the diffraction pattern of tetragonal (I4/mcm space group) $MAPbI_3$ collected at 0.004 GPa and 0.060 GPa and room temperature. The external compression shifts the diffraction peaks to higher 2θ values, while the crystal phase remains constant in the range of pressures explored. **Figure 1b** shows the diffraction pattern of $MAPbI_3$ collected at 333 K. At this temperature, $MAPbI_3$ is in a cubic crystal structure (Pm-3m space group). Also in this case, we observe the shift of the reflections towards higher 2θ values without indication of phase transition induced by external pressure. In **Figure 1c–d**, we plot the lattice parameters of tetragonal $MAPbI_3$ (room temperature, where $a = b \neq c$, **Figure 1c**), and cubic $MAPbI_3$ (333 K, where $a = b = c$, **Figure 1d**), as a function of external pressure. This pressure-induced variation of the lattice parameters was determined from refinements using the Rietveld method, *i.e.*, the obtained diffractograms were fitted using Pseudo–Voigt functions considering the unit cell dimensions and atomic coordinates as variables. Further details on the method are provided in the **Method Section** of the **Supporting Information**. In all cases, as expected, applying external pressure leads to a decrease in lattice parameter.

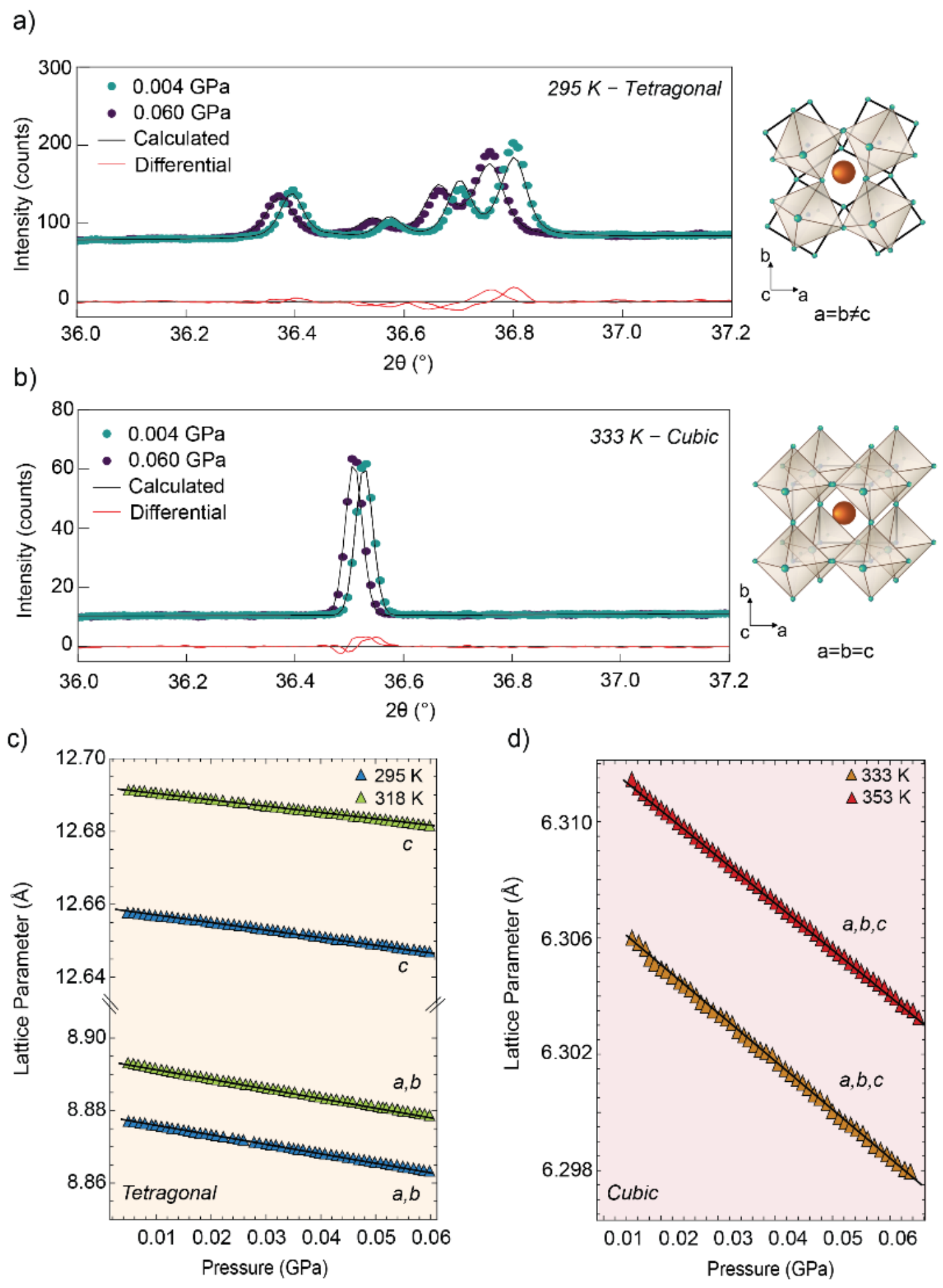

**Figure 1. a)** Selected reflections of the diffraction pattern of $MAPbI_3$ collected at room temperature and **b)** at 333 K, at 0.004 (green) and 0.060 GPa (purple), together with schematic drawings of the crystal structures. In black, the calculated profile structure and in red the residuals of the fit. Pressure-dependent lattice parameters of **c)** tetragonal $MAPbI_3$ (I4/mcm space group)[33] collected at 295 and 318 K, and **d)** cubic $MAPbI_3$ (Pm-3m space group) collected at 335 and 355 K obtained from refinements using the Rietveld method. Fits are shown with solid black lines.

In order to investigate the role of the halide on the temperature-dependent elastic properties, we have performed a similar analysis on the mixed-halide perovskites $MAPb(Cl_{1-x}Br_x)_3$ and $MAPb(I_{1-x}Br_x)_3$, as reported in **Figure S1–S7**. Except for $MAPbI_3$, which has a tetragonal (I4/mcm space group) symmetry in the range of 298 and 330 K,[34,35] these perovskites show cubic symmetry (Pm-3m space group) under these conditions.[22,36] For all single and mixed-halide perovskites, we observed a linear decrease of the volume with pressure for the temperatures explored. The isothermal equation of state for a solid is given by the bulk modulus, B:

$$B = \frac{pressure}{strain} = -\frac{\Delta P}{\Delta V}V \quad (1)$$

with $\Delta P/\Delta V$ the derivative of pressure with respect to volume, and $V$ the volume at ambient pressure. By fitting this function to the pressure–volume trends (**Figure S1–S7**), we estimated the bulk moduli (B) of the single halide (iodide, bromide, chloride) and mixed-halide series $MAPb(Cl_{1-x}Br_x)_3$ and $MAPb(I_{1-x}Br_x)_3$. The results are shown in **Figure 2a** and reported in **Table S1** in **Supporting Note 1**. In the entire temperature range of 298 – 355 K, $MAPbBr_3$ (halide fraction $x = 1$) has a bulk modulus of 17 GPa, which gradually increases on replacing the $Br^-$ with $Cl^-$. An increase of the bulk modulus is associated with an increase in the stiffness of the material. At room temperature, $MAPbCl_3$ has a bulk modulus of 20 GPa, which seems to be slightly lower (19 GPa) at elevated temperatures. However, over the entire temperature range, the bromide-based perovskites are significantly softer than the perovskites containing chloride. The introduction of iodide leads to further softening of the perovskites, gradually decreasing the bulk modulus to 14 GPa. This softening of the bulk modulus with larger halides is consistent with previous single crystal studies reporting smaller Young's modulus of the {100} crystal facet.[37]

For the cubic systems, we observe a linear relation between the bulk modulus and the average halide radius as common for metal alloys (**Figure S8**).[38] Such findings confirm previous assumptions on similar compositions,[20] but are in contrast with theoretical predictions for $CsPb(I_{1-x}Br_x)_3$.[39] In order to investigate the role of the $MA^+$ cation in the halide perovskite and the role of the trivalent metal in the elpasolite structure on their elastic properties, we determined the temperature-dependent bulk moduli of $CsPbBr_3$ and -$Cl_3$, as well as bulk moduli of $Cs_2AgBiBr_6$ and -$Cl_6$, and $Cs_2AgInCl_6$. The results are shown in **Figures 2b–c** and **Figure S9–13** while numeric values are reported in **Table S1** in **Supporting Note 1**. Even though the Cs-based perovskites have different crystal structures than the MA-based ones, the room temperature bulk moduli of $CsPbBr_3$ (B = 17.5 GPa, see **Figure S12**) and -$Cl_3$ (B = 19 Gpa, see **Figure 2c** and **Figure S13**) are almost identical to their MA-based counterparts (*i.e.,* less than 1 Gpa variation). These observations suggest that the mechanical properties are mostly defined by the framework of corner-sharing $PbI_3$ ($Cl_3$, $Br_3$) octahedra. Regarding the B-site cation, we observe that replacing $Pb^{2+}$ with both $Ag^+$ and a trivalent cation ($Bi^{3+}$ or $In^{3+}$) leads to a stiffening of the lattice, associated with larger bulk moduli, *i.e.,* 22-23 Gpa for the ones with $Cl^-$ and 19 Gpa for $Cs_2AgBiBr_6$.[26] However, the effect of changing the cations is relatively small compared to the effect of changing the halides. Hence, we can conclude that the halide framework dominates the mechanical properties of halide perovskites and elpasolites, rather than the A- or B-site cations. Another general finding is that most of the bulk moduli only show slight changes with temperature in the range from 298 to 355 K, in contrast with the expected decrease of B with temperature observed in oxide perovskites, and the expected increase calculated for tin-based halide perovskites such as $CsSnI_3$.[40] As can be seen **in Figures 2a** and **Figure 2b**, the variation of the bulk modulus with the halide is larger than any temperature effect in this range. Two notable exceptions include $CsPbCl_3$ and $MAPbI_3$, that undergo temperature-dependent phase transitions in between 298 and 330 K, shown in **Figure 2c**. Our data hint towards a softening of the materials (*i.e.,* lower bulk modulus) close to their transition temperature (dotted lines). This finding is particularly relevant since temperature-dependent elastic properties may introduce stress during fabrication of halide perovskite and elpasolite thin films or during temperature cycling processes in operating devices.

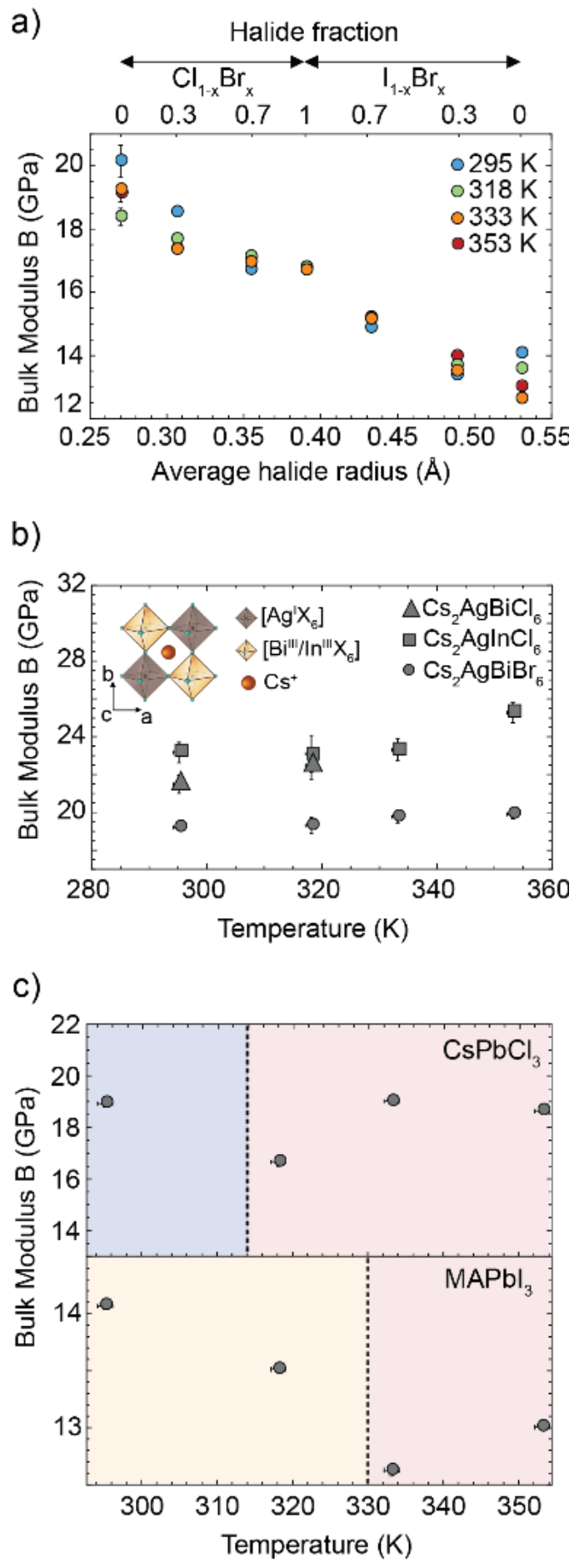

**Figure 2**. **a)** Bulk modulus (B) of methylammonium (MA) lead mixed-halide perovskites as a function of the average halide radius and the mixing halide (Cl: left, Br: middle, I: right) ratio at 295, 318, 335, 355 K. This shows that the Cl-based materials soften on the addition of $Br^-$, and that $I^-$ leads to further softening of the lattice. Note that, except for $MAPbI_3$, all these perovskites exhibit a cubic symmetry, and their bulk modulus is roughly constant with temperature. **B)** Bulk modulus of halide double perovskites $Cs_2AgBiBr_6$ (circle), $Cs_2AgInBr_6$ (square), and $Cs_2AgBiCl_6$ (triangle) as a function of temperature. Also here, the Cl-based compounds are stiffer than the $Cs_2AgBiBr_6$, with only minor temperature effects. **C)** Bulk modulus of $MAPbI_3$ and $CsPbCl_3$ as a function of temperature. The regions in blue, yellow, and red correspond to the orthorhombic, tetragonal, and cubic phase, respectively. The dashed lines correspond to the temperature at which the phase transition occurs.

Considering that the bulk modulus is a volume property, while the elastic properties of non-cubic perovskites, such as tetragonal $MAPbI_3$, are in general anisotropic and therefore depend on crystallographic direction, we derived the compressibility of the specific crystal axes by monitoring their compression as a function of applied pressure (*e.g.*, $K_a$, for the compressibility along a-axis):

$$K_a = -\frac{(a_f - a_0)}{a_f} P \tag{2}$$

with $a$ the lattice parameter, and $a_f - a_0$ the change in lattice parameter between the lattice parameter at the final ($a_f$) and initial ($a_0$) pressure. Hence, we determined the compressibility for *a*-, *b*-, and *c*-axis for non-cubic perovskites. The volumetric compressibility, $K_V$, *i.e.*, the susceptibility of a material to compress upon external applied pressure, is inversely proportional to B and is calculated as reported in **Supporting Note 2** while values are reported in **Table S5**. The compressibility along the *a*-, *b*- and *c*-axis for the compositions studied are reported in **Table S2–4**. **Figure 3** shows that compression along the *a*- and *b*- axes of tetragonal $MAPbI_3$ and orthorhombic $CsPbCl_3$ is significantly larger than for the *c*-axis. In addition, we observe that the compressibility of *c* increases significantly in $CsPbCl_3$ during the orthorhombic-to-cubic phase transition as well as in $MAPbI_3$ during the tetragonal-to-cubic phase transition, as it becomes identical to the *a*- and *b*-axes. In the orthorhombic $CsPbBr_3$, we find that compression along the *a*- and *c*-axes is significantly larger than for the *b*-axis. Additional pressure studies using an extended temperature range are needed to determine the origin of the observed axis-dependent compressibility. For $CsPbBr_3$, we observe a gradual increase of the compressibility of the *b*- and *c* axis with temperature, in combination with swapping of the *a*- and *b*-axis compressibility at 353 K. This phenomenon may occur in the case of an anti-isostructural phase transition, although more measurements are needed to conclusively show whether such a transition exists in $CsPbBr_3$. In order to investigate the relation between pressure response and thermal expansivity, we further measured the temperature-dependent XRD of single-halide perovskites ($MAPbI_3$, $MAPbBr_3$, $MAPbCl_3$, $CsPbBr_3$, $CsPbCl_3$) and elpasolites ($Cs_2AgBiBr_6$, $Cs_2AgBiCl_6$, $Cs_2AgInCl_6$) between 100 and 400 K. A similar Rietveld refinement was used to study the temperature dependence of the crystal axes *a*, *b* and *c*. The results are shown in **Figure 4**. Here, blue areas represent orthorhombic (Pnmb space group), yellow areas represent tetragonal (I4/mcm space group), and red areas represent cubic symmetry (Pm-3m space group). Although similar experiments have been reported for a selection of these compositions,[34,36,41,42] we here compare thermal expansivity of all compositions using identical synthesis and measurement conditions and link the thermal response to the elastic properties of the same materials.

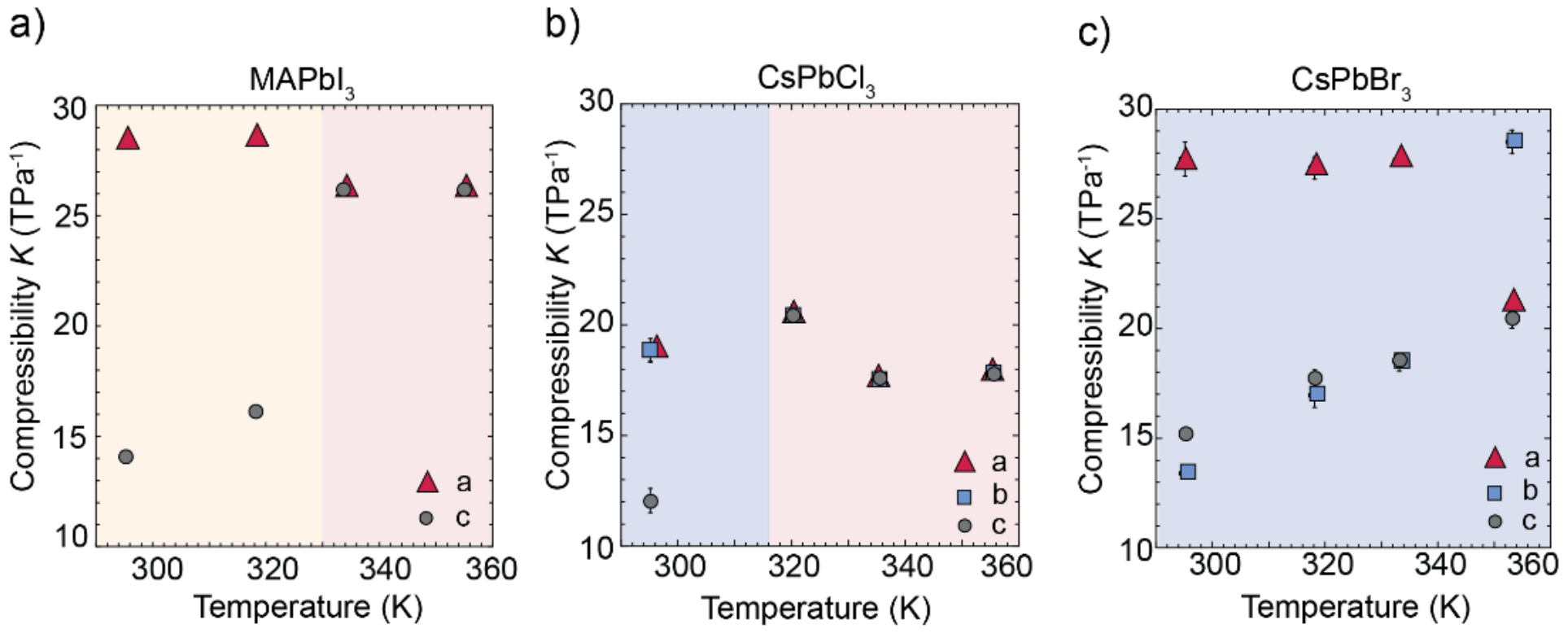

**Figure 3. a)** Compressibility $K$ of the crystal axes in the tetragonal (yellow) and cubic (red) phase of $MAPbI_3$. The compressibility of the *c*-axis is smaller compared to the *a*-axis in the tetragonal phase. **b)** Compressibility $K$ of the crystal axes in the orthorhombic (blue) and cubic (red) phase of $CsPbCl_3$. The compressibility of the *c*-axis is smaller compared to the *a*-axis and *b*-axis in the orthorhombic phase. **c)** Compressibility $K$ of the crystal axes in the orthorhombic (blue) phase. The compressibility of the *b*-axis is smaller compared to the *a*- and *c*- axis in the tetragonal phase up to 353 K where an abrupt change is observed. At 353 K, the compressibility of the *b*-axis is larger than the compressibility of *a*- and *c*-axis. Error bars are reported in the figure and are in some cases smaller than the data point size.

Comparing the different halide perovskites and elpasolites in cubic symmetry (red areas), we observe an increase in lattice parameters as a function of temperature, associated with a positive thermal expansivityas shown in **Table 1**. In line with the halide-dependent trend in bulk modulus, for all compositions studied here the thermal expansivity is observed to increase following Cl<Br<I, due to the softness of the material. We observe that the elpasolites are less expandable with temperature than the lead-halide perovskites, with thermal expansivity of $2.86 \times 10^{-5}$ $K^{-1}$ for $Cs_2AgBiBr_6$ and $2.42 \times 10^{-5}$ $K^{-1}$ for $Cs_2AgBiCl_6$. This observation is consistent with their larger bulk moduli, showing that the elpasolite structure is more resistant to both temperature and pressure changes. Furthermore, the absolute values are close to recent theoretical predictions.[43]

| | | Orthorhombic ($\times 10^{-5}$ K$^{-1}$) | Tetragonal ($\times 10^{-5}$ K$^{-1}$) | | Cubic ($\times 10^{-5}$ K$^{-1}$) |
|---|---|---|---|---|---|
| **$MAPbI_3$** | a | 3.81 | a=b | Function* | 4.04 |
| | b | 2.02 | | | |
| | c | 5.81 | c | 0.34 | |
| **$MAPbBr_3$** | a | 9.04 | a=b | 6.99 | 3.53 |
| | b | −2.79 | | | |
| | c | 1.00 | c | Function* | |
| **$MAPbCl_3$** | a | Function* | - | | 3.51 |
| | b | Function* | | | |
| | c | Function* | | | |
| **$CsPbBr_3$** | a | 7.68 | - | | 3.27** |
| | b | Function* | | | |
| | c | 3.51 | | | |
| **$CsPbCl_3$** | a | Function* | - | | 3.02 |
| | b | Function* | | | |
| | c | Function* | | | |
| **$Cs_2AgBi(Br_{0.33}I_{0.67})_6$** | | - | - | | 5.08** |
| **$Cs_2AgBiBr_6$** | | - | - | | 2.86 |
| **$Cs_2AgBiCl_6$** | | - | - | | 2.42 |
| **$Cs_2AgInCl_6$** | | - | - | | 2.49 |
| **$Cs_2Ag(Bi_{0.5}In_{0.5})Br_6$** | | - | - | | 2.84** |
| **$Cs_2Ag(Bi_{0.5}Sb_{0.5})Br_6$** | | - | - | | 3.34** |
| **$Cs_2Ag(Bi_{0.9}Fe_{0.1})Br_6$** | | - | - | | 3.16** |

*Non-linear expansivity (see **Supporting Note 3**).
**Data collected with a lab-based X-ray diffractometer.

**Table 1.** Thermal expansivity along all crystal axes for all compositions. If the temperature response is non-linear the experimental data is fitted using a second order polynomial and denoted by 'Function'. The optimized parameters of these fits are given in the **Supporting Information** (see **Supporting Note 3**). As the temperature range in which the materials are orthorhombic, tetragonal, or cubic varies with composition, the reported (functions of) thermal expansivity are only valid in that specific temperature range.

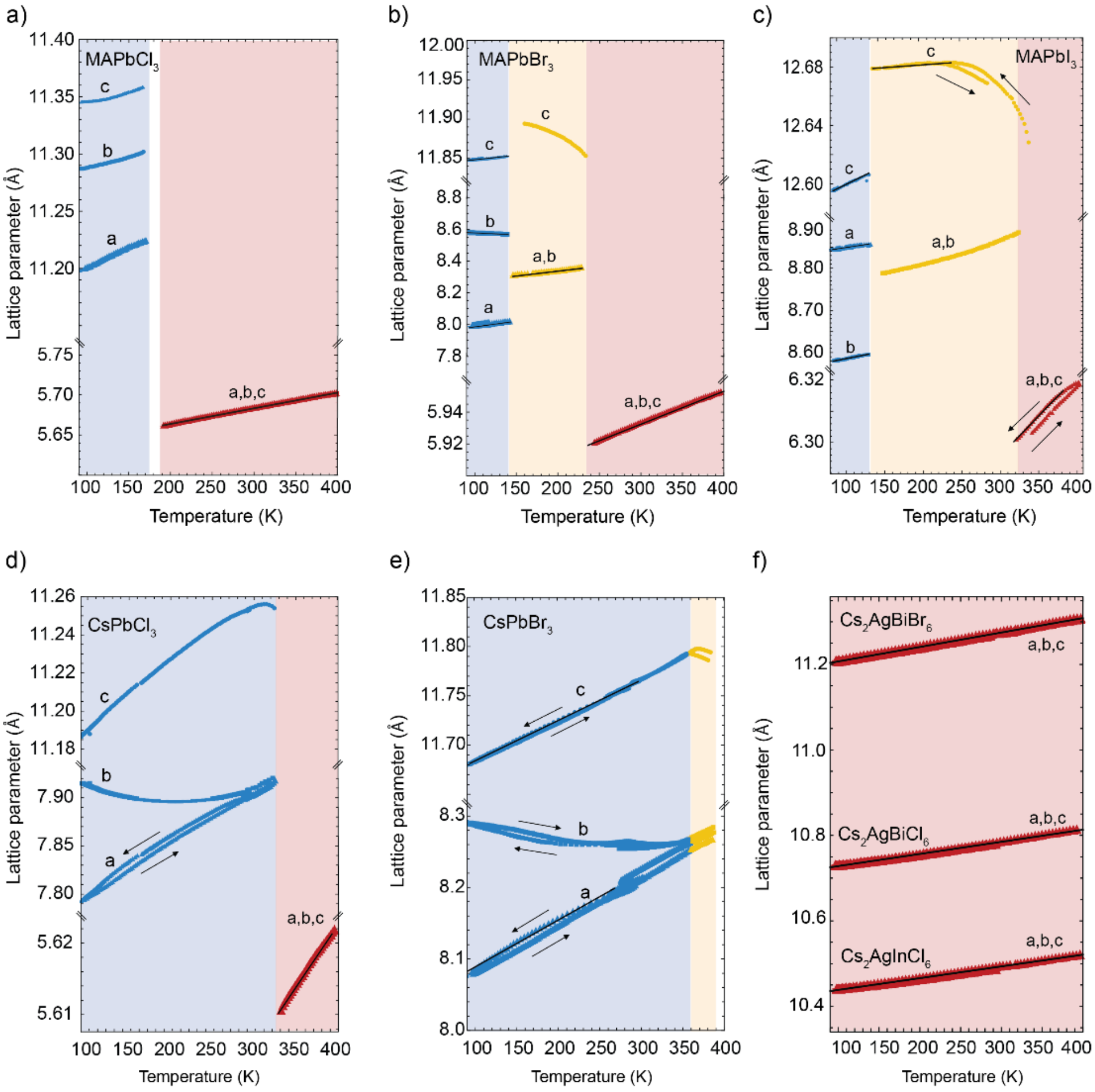

**Figure 4.** Temperature-dependent lattice parameters of **a)** $MAPbCl_3$, **b)** $MAPbBr_3$, **c)** $MAPbI_3$, **d)** $CsPbCl_3$, **e)** $CsPbBr_3$ and **f)** $Cs_2AgBiBr_6$, $Cs_2AgBiCl_6$, $Cs_2AgInCl_6$ obtained from profile Rietveld refinement of synchrotron XRD patterns collected between 90 K and 400 K. The blue, yellow and red regions correspond to the orthorhombic, tetragonal and cubic phase, respectively. The regions in white correspond to the temperature range where two phases coexist in the XRD patterns. The black arrows indicate the direction of temperature variation, *i.e.* cooling (arrow down) and heating (arrow up). Solid lines represent the fit for the linear thermal expansivity.

Additional lab-based temperature-dependent XRD on elpasolite with mixed trivalent metals, *i.e.,* $Cs_2Ag(Bi_{0.5}Sb_{0.5})Br_6$, $Cs_2Ag(Bi_{0.5}In_{0.5})Br_6$, $Cs_2Ag(Bi_{0.9}Fe_{0.1})Br_6$, and mixed-halide such as $Cs_2AgBi(Br_{0.33}I_{0.67})_6$, (**Supporting Note 4** and **Figure S14**) indicate higher thermal expansivity for elpasolites containing $I^-$ ions, identical to the lead-based perovskites (see **Table 1**). Previous work already showed a minor role of the Cs cation on the thermal expansivity in halide elpasolites,[44] and we here find that also the metal cation does not affect the thermal expansivity. For phases with lower symmetry than cubic, *i.e.*, orthorhombic (blue) and tetragonal (yellow), thermal expansivity is anisotropic. Interestingly, the axes-dependent thermal expansivity and compressibility in tetragonal crystal structure follow the same trend (*i.e.*, *c*-axis is less expandable and more compressible compared to axis *a* and *b*). On the other hand, in the orthorhombic crystal phase we observe no such relation suggesting that thermal expansivity and compressibility are decoupled in this phase. An additional consideration is that some lattice parameters have non-linear temperature dependent thermal expansivity. For example, the axes of orthorhombic $CsPbBr_3$ and $CsPbCl_3$ and of tetragonal $MAPbI_3$, show a non-linear dependence of lattice parameter with temperature, so that the thermal expansivity becomes temperature-dependent (see **Supporting Note 3** and **Table S6**). Notably, in all halide perovskites studied here, the *c*-axis of the tetragonal phase has a negative thermal expansivity in at least a part of the temperature range. According to Landau theory such a behavior reflects bi-linear coupling of the order parameter and spontaneous lattice strain.[45] Considering that negative thermal expansivity values (*i.e.,* thermal contraction) are mainly observed at temperatures 'close' to thermal phase transitions, it seems likely that these two phenomena are related. Previous work has reported similar behavior for $MAPbI_3$ and formamidinium lead iodide ($FAPbI_3$) perovskites[46–48] as well as several oxide-based materials that undergo phase transitions.[49] For $CsPbBr_3$ and $CsPbCl_3$, the temperature range with negative thermal expansivity for the *c*-axis is relatively small (20–40 K). In contrast, for $MAPbBr_3$ and $MAPbI_3$, the negative thermal expansivity of the *c*-axis is already observed 100 K below the tetragonal-cubic phase transition. This observation, together with temperature-dependent bulk modulus from **Figure 2a**, shows that the range in which the phase transition is affecting thermal and elastic properties is substantially broader than previously thought on basis of the (much smaller) temperature range in which crystal phases co-exist.[34] This has major implications for synthesizing thin films on substrates with different thermal expansivity than those exhibited by the perovskite, where the temperature gradient during cooling from annealing to room temperature will introduce strain in the material.

## Conclusions

We determined temperature-dependent elastic properties of several halide perovskites (namely single- and mixed-halide) and elpasolite (namely mixed halide and mixed trivalent metals) from pressure- and temperature-dependent diffraction experiments using a synchrotron radiation source. We used temperatures ranging from 298 to 363 K (*i.e.*, ambient to 90°C), which is a typical temperature range for thin film synthesis and photovoltaic operation conditions. To obtain information relevant to ambient phases, we explored pressures from ambient to 0.06 GPa. We find that the elastic properties and the thermal expansivity in both halide perovskites and elpasolites are dominated by the halide framework, with lower bulk moduli for larger halide radius size (Cl<Br<I). Contrary, the monovalent cation and the trivalent metal have a minor role in determining elastic properties. The bulk modulus of these materials remains constant within the investigated temperature range, provided that the crystal structure remains unchanged. On the other hand, we observe that thermal phase transitions lead to a decrease in the bulk modulus (*i.e.*, softening of the material), as exemplified by $MAPbI_3$ and $CsPbCl_3$. For non-cubic systems, where the elastic properties are anisotropic, we obtained the axes-dependent compressibility. Our results suggest that the *c* axis is much softer than the *a*- and *b* axes in the tetragonal $MAPbI_3$ and orthorhombic $CsPbCl_3$, whereas the *b*-axis is softer in $CsPbBr_3$. Furthermore, we find that some temperature-dependent phase transitions of halide perovskites are accompanied by negative thermal expansivity of certain crystal axes, already several tens of degree below the phase transition temperature. This observation shows that the range in which the phase transition is affecting thermal and elastic properties is substantially broader than previously thought on basis of the (much smaller) temperature range in which phases co-exist. Altogether, these findings have major implications for synthesizing thin films at substrates with different thermal expansivities than the perovskite and elpasolite, where the temperature gradient during the synthesis and the temperature cycling during device operation will introduce strain in the material. Hence, acquiring a deep comprehension of the temperature-dependent elastic properties in these materials is essential to bolster the advancements in strain engineering which has proven to be a potent tool for manipulating both the optical properties and the stability of perovskite thin films.

**Supporting Information**

Experimental details, materials, and methods; lattice parameters as a function of pressure, P-V curves, temperature-dependent bulk moduli and axes-dependent compressibility for all the compositions explored; *in-situ* temperature-dependent XRD of several elpasolites; tables including temperature-dependent bulk moduli, axes-dependent compressibility and temperature-dependent expansivity; lab-based temperature dependent X-ray diffraction for additional elpasolites; calculation of volumetric and axes-dependent thermal expansivity and compressibility; calculation of the bulk moduli. Data and fit procedures reported in this study can be accessed at https://doi.org/XXX and are available under a CC-BY Creative Commons Attribution 4.0 International license.

**Acknowledgments**

E.M.H. thanks the Dutch Research Council for funding under the grant number VI.Veni.192.034. L.A.M. acknowledges funding from the Dutch Research Council (NWO) under the grant number OCENW.XS22.2.039. B.B. acknowledges funding by the Austrian Science Fund (FWF) under the project number J4607-N. Prof. Bert Weckhuysen is gratefully acknowledged for facilitating the supporting *in-situ* XRD measurements. The authors acknowledge the European Synchrotron Radiation Facility (ESRF) for provision of synchrotron radiation beamtime at the Swiss-Norwegian beamline BM01 (proposal MA5378). The authors acknowledge funding from the Advanced Research Center Chemical Building Blocks Consortium (ARC CBBC).

# Supporting Information to

## Which Ion Dominates Temperature and Pressure Response of Halide Perovskites and Elpasolites?

Loreta A. Muscarella,[1#*] Huygen J. Jöbsis,[1#] Bettina Baumgartner,[1] P. Tim Prins,[1] D. Nicolette Maaskant,[1] Andrei V. Petukhov,[2] Dmitry Chernyshov,[3] Charles J. McMonagle,[3] and Eline M. Hutter[1*]

1. Inorganic Chemistry and Catalysis group, Debye Institute for Nanomaterials Science and Institute for Sustainable and Circular Chemistry, Department of Chemistry, Utrecht University, Princetonlaan 8, 3584 CB Utrecht, the Netherlands
2. Physical and Colloid Chemistry, Debye Institute for Nanomaterials Science, Department of Chemistry, Utrecht University, Padualaan 8, 3584 CH Utrecht, the Netherlands
3. Swiss–Norwegian Beamlines, European Synchrotron Radiation Facility, 71 Avenue des Martyrs, 38000 Grenoble, France

# These authors contributed equally

* Correspondence should be addressed to E.M.Hutter@uu.nl and loretaangela.muscarella@gmail.com

## Experimental Section

### Materials

The perovskite and elpasolite powders are prepared using lead (II) iodide ($PbI_2$; TCI, purity 99.99%, trace metals basis), methylamine hydroiodide (MAI; TCI, purity >99%), lead (II) bromide (TCI, purity 98%), and methylamine hydrobromide (MABr; purity TCI, >98%), lead (II) chloride ($PbCl_2$; TCI, purity >99%) and methylamine hydrochloride (MACl; TCI, purity >98%), cesium iodide (CsI; TCI, purity > 99%), cesium bromide (CsBr; TCI, purity > 99%), cesium chloride (CsCl; TCI, purity > 99%), silver (I) chloride (AgCl; Merck, purity 99.999% trace metals basis), silver (I) bromide (AgBr; Alfa Aesar, Premion®, purity 99.998% metals basis), bismuth (III) iodide ($BiI_3$, Merck, purity 99%), bismuth (III) bromide ($BiBr_3$; Merck, purity ≥ 98%), bismuth (III) chloride ($BiCl_3$; Merck, purity 99.998% trace metals basis), indium (III) bromide ($InBr_3$, Alfa Aesar, purity 99.99% (metal basis)), indium (III) chloride ($InCl_3$; TCI, purity >99.0%) and, antimony (III) bromide ($SbBr_3$, Alfa Aesar, purity ultra-dry 99.999% (metal basis)), iron (III) bromide ($FeBr_3$, Alfa Aesar, purity >98%). The precursor powders are obtained weighing the components with the desired molar stoichiometric ratio without further purification.

### Perovskite powder preparation

Methylammonium-based perovskite powders are prepared using mechanochemical synthesis by grinding the precursor salts in a ball mill (Retsch Ball Mill MM-400) using a grinding jar of 10 ml and two stainless steel balls (⌀10 mm) for 60 minutes at 30 Hz. $CsPbX_3$ (X= I, Br, Cl) and elpasolite powders are prepared milling at 30 Hz for 90 minutes. No annealing is performed after the mechanochemical synthesis.

### Pressure and temperature dependent powder X-ray diffraction

Pressure and temperature dependent powder X-ray diffraction data were collected at the Swiss–Norwegian Beamline BM01 (ESRF, Grenoble), with a $\lambda = 0.9590$ Å.[1] The powdered samples were loaded into a sapphire capillary pressure cell with a 1 mm external and 0.6 mm internal diameter (CRYTUR, spol. S r.o., Czech Republic).[2] The pressure cell was filled with a fluorinated inert liquid (FC-770, 3M) as a pressure-transmitting medium and the pressure was swept from 0.004 to 0.060 GPa at temperatures of 298, 318, 335, and 355 K. Temperature control is obtained with an Oxford Cryosystems cooler,[3] where the temperature offset caused by the large diameter sapphire capillary was calibrated using the diffraction pattern of silver.[4] The Dectris Pilatus 2 M detector was used for recording 2D diffraction images and the local program Bubble was used for integration of the 2D images.[1]

**Temperature-dependent, lab-based X-Ray diffraction**

The additional temperature-dependent X-ray diffractograms were collected using a Bruker Axs D8 Phaser Advanced. Equipped with a Cu $K_{\alpha1,2}$ ($\lambda$ = 1.54184 Å) radiation source operating at 40 kV and 40 mA and an Anton Paar XRK900 Temperature Chamber. The sample height for all experiments was aligned so that the (400) reflection of $Cs_2AgBiBr_6$ was detected at 31.7°. The temperature was calibrated by measuring the lattice expansivity of MgO. For $CsPbBr_3$ the additional diffractograms were collected from 330 K to 500 K (and back) with a temperature increment of 1 K. The additional elpasolite compositions were studied between 325 K and 450 K as discussed in See **Supporting Note 4**.

**Rietveld Refinement method**

The refinements using the Rietveld method was performed using the FullProfSuite software. All diffractograms were fitted using Pseudo–Voigt functions with the unit cell dimensions, reflection intensity, atomic coordinates of the inorganic elements as variables. In some cases the Debye–Waller factor and the profile shape were used as variables to improve the quality of the fits. Note that including these variables only lower the residuals, without changing the lattice parameters obtained from the fits. The starting crystallographic information are obtained using files (CIF) retrieved from the ICSD database. Below a list of the used entries is given.

| **Composition, phase(s)** | **ICSD entry/entries** |
|---|---|
| $MAPbI_3$, orthorhombic, tetragonal, cubic | 428898, 238610, 7236651 |
| $MAPbBr_3$, orthorhombic, tetragonal, cubic | 268782, 268779, 268785 |
| $MAPbCl_3$, orthorhombic, tetragonal, cubic | 1469989**, 7236651 |
| $MAPb(Br_{0.30}I_{0.70})_3$ | 243598** |
| $MAPb(Br_{0.70}I_{0.30})_3$ | 243598** |
| $MAPb(Cl_{0.30}I_{0.70})_3$ | 243598** |
| $MAPb(Cl_{0.70}I_{0.30})_3$ | 243598** |
| $CsPbBr_3$, orthorhombic, tetragonal, cubic | 14608, 14610 |
| $CsPbCl_3$, orthorhombic, cubic | 230496, 249888, 29072 |
| $Cs_2AgBi(Br_{0.33}I_{0.67})_6$, cubic | 252164** |
| $Cs_2AgBiBr_6$, cubic | 252164 |
| $Cs_2AgBiCl_6$, cubic | 252451 |
| $Cs_2AgInCl_6$, cubic | 257115 |
| $Cs_2Ag(Bi_{0.5}In_{0.5})Br_6$, cubic | 252164** |
| $Cs_2Ag(Bi_{0.5}Sb_{0.5})Br_6$, cubic | 252164** |
| $Cs_2Ag(Bi_{0.9}Fe_{0.1})Br_6$, cubic | 252164** |

** For these compositions, no structural files (.cif) were available. In these cases we have used cif's of identical structures and space groups to refine the lattice parameters. For $MAPbCl_3$, we have utilized the structural file of cubic $MAPbI_3$ (ICSD 7236651). For the MA-based mixed halide compositions, we have used the structural file of $MAPb(Br_{0.15}I_{0.85})_3$ (ICSD 243598). For elpasolites with mixed trivalent cations, we have utilized the structural file of cubic $Cs_2AgBiBr_6$ (ICSD 252164). Here, we assumed stoichiometric occupancy on the metal site, which may introduce some discrepancies in fitting the relative peak heights.

## Supporting Figures

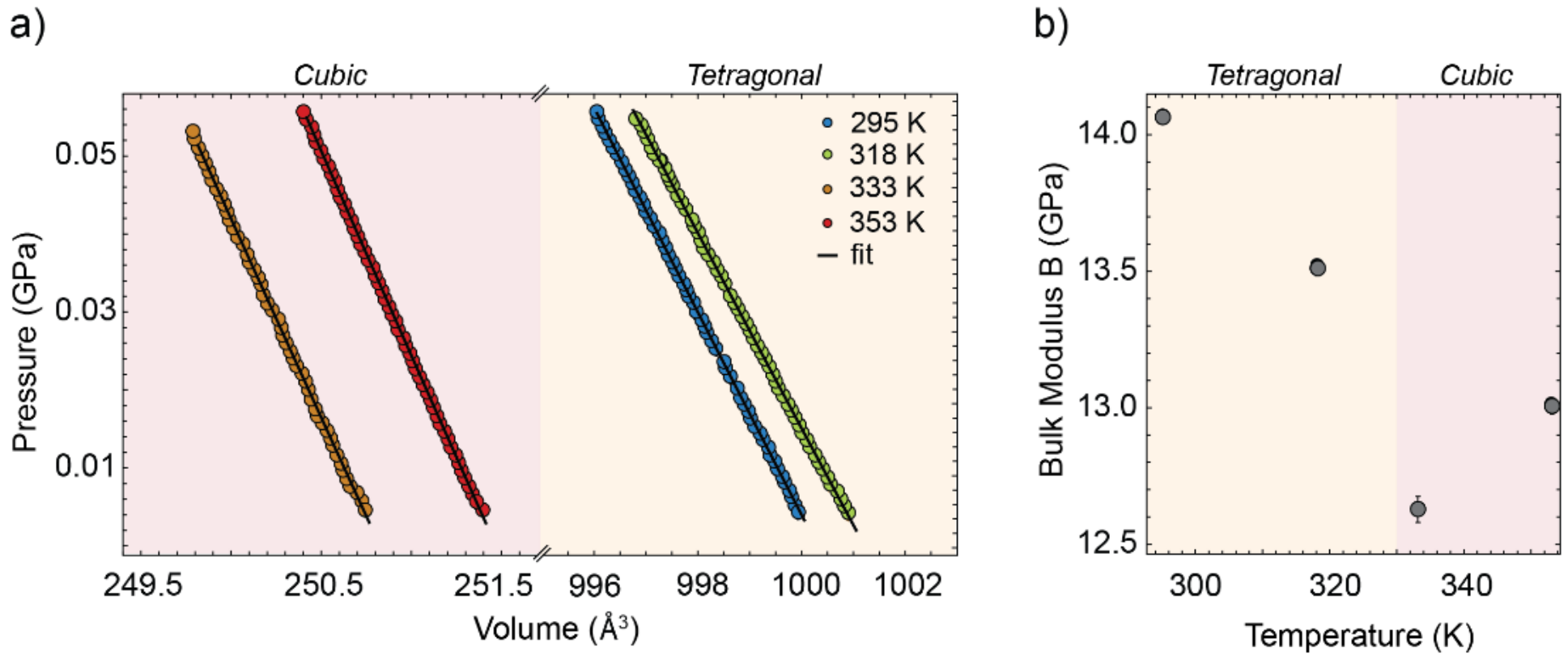

**Figure S1. a)** Unit-cell volume changes with pressure for $MAPbI_3$ in the tetragonal I4/mcm (yellow) and cubic Pm-3m (red) phase and the corresponding fit to equation 1 of the main text. **b)** Bulk moduli obtained from the data shown in (a) and reported as a function of temperature in the tetragonal (yellow) and cubic (red) phase. The lattice parameters versus pressure are shown in **Figure 1** of the main text.

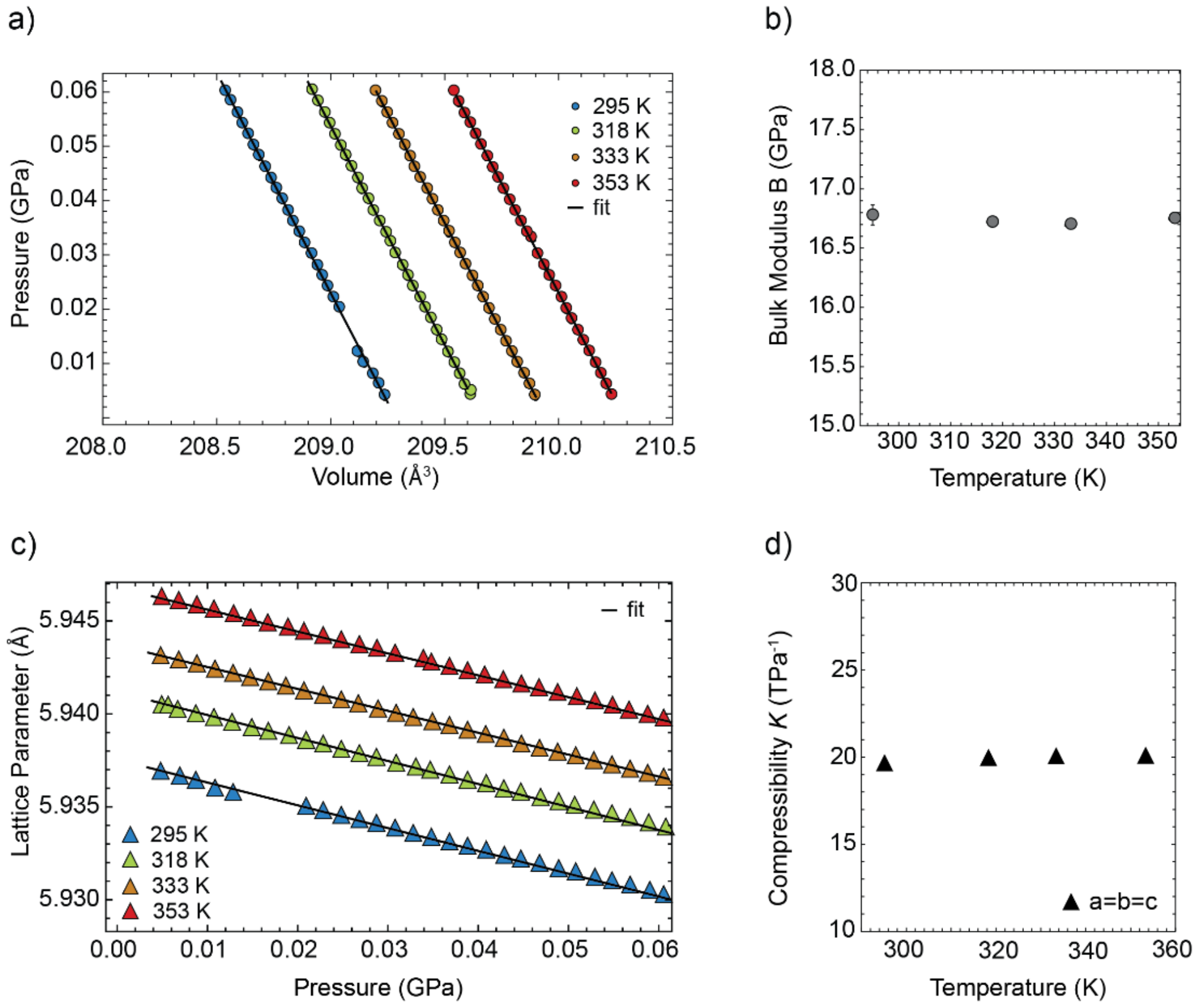

**Figure S2.** Unit-cell volume changes with pressure for $MAPbBr_3$ in the cubic Pm-3m phase and the corresponding fit to equation 1 of the main text. **b)** Bulk moduli obtained from the data shown in (a) and reported as a function of temperature in the cubic phase show no temperature dependence. **c)** Pressure-dependent lattice parameters of cubic $MAPbBr_3$ (Pm-3m space group) collected at 295, 318, 333, and 353 K obtained from refinements of synchrotron XRD pattern using the Rietveld method. **d)** Compressibility *K* of the lattice parameters in the cubic phase. Given the cubic structure, the compressibility is equal along all three axes.

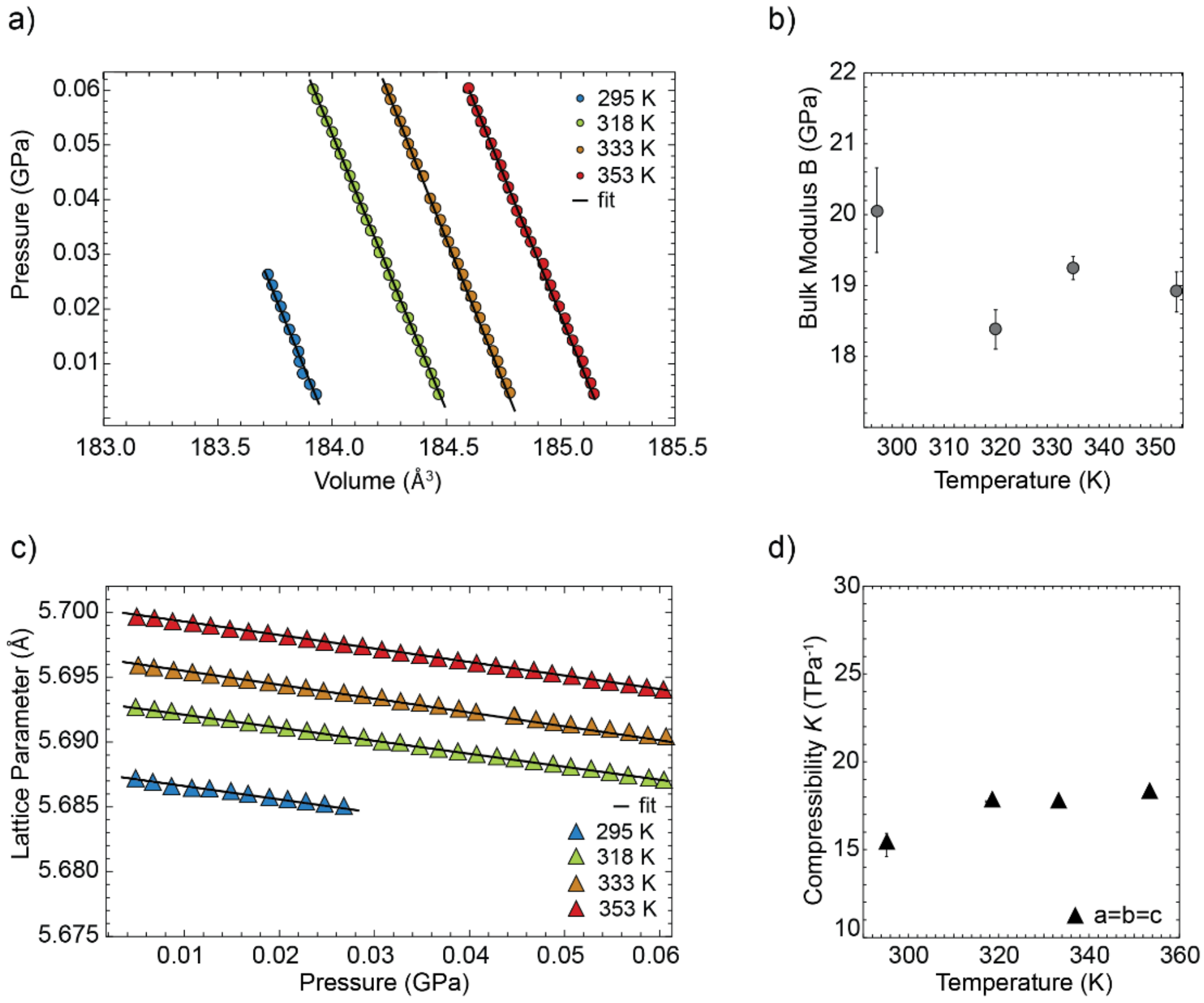

**Figure S3.** Unit-cell volume changes with pressure for $MAPbCl_3$ in the cubic Pm-3m phase and the corresponding fit to equation 1 of the main text. **b)** Bulk moduli obtained from the data shown in (a) and reported as a function of temperature in the cubic phase show no temperature dependence. **c)** Pressure-dependent lattice parameters of cubic $MAPbCl_3$ (Pm-3m space group) collected at 295, 318, 333, and 353 K obtained from refinements of synchrotron XRD pattern using the Rietveld method. **d)** Compressibility *K* of the lattice parameters in the cubic phase. Given the cubic structure, the compressibility is equal along all three axes.

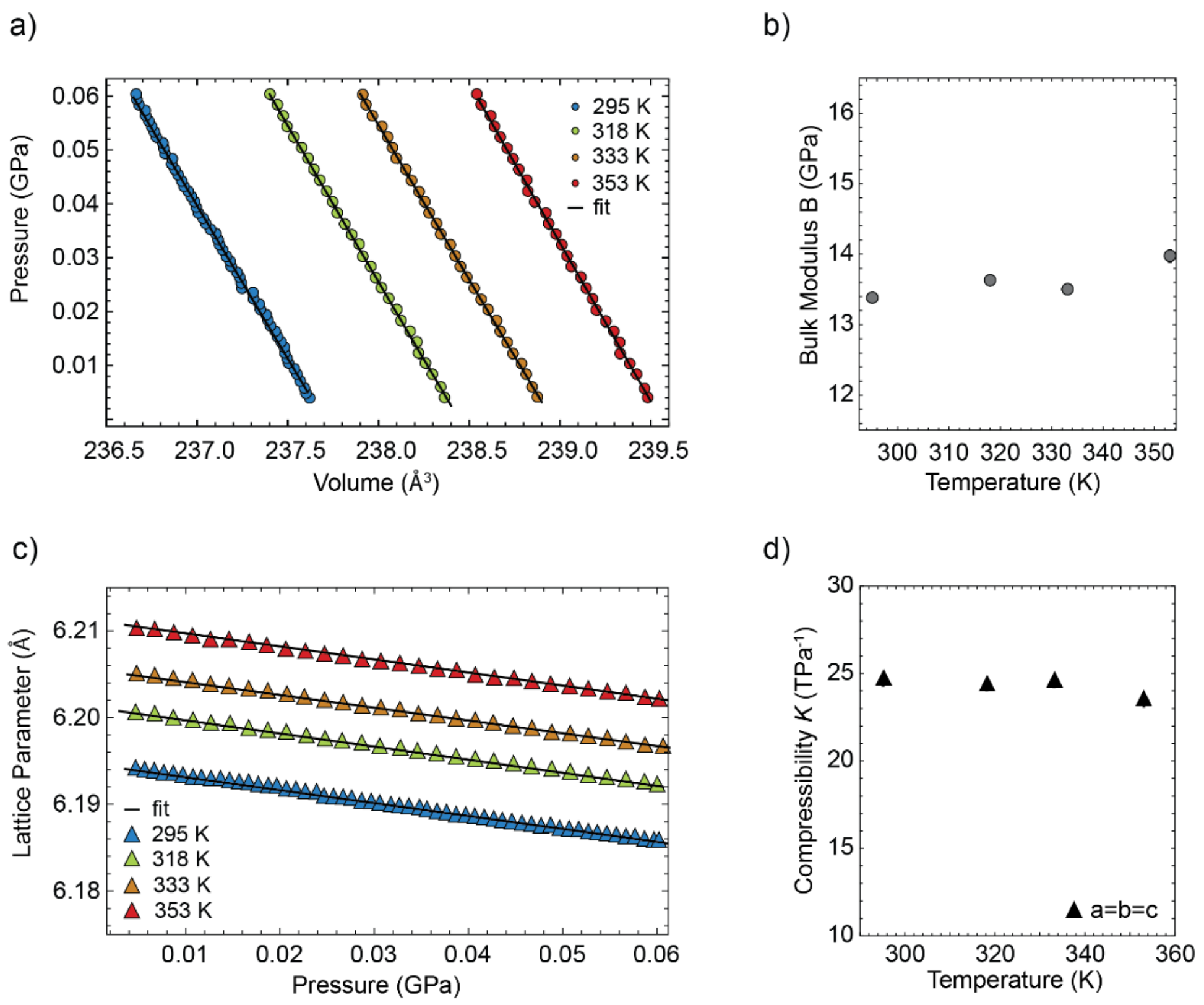

**Figure S4**. Unit-cell volume changes with pressure for MAPb($Br_{0.3}I_{0.7}$)$_3$ in the cubic Pm-3m phase and the corresponding fit to equation 1 of the main text. **b)** Bulk moduli obtained from the data shown in (a) and reported as a function of temperature in the cubic phase show no temperature dependence. **c)** Pressure-dependent lattice parameters of cubic MAPb($Br_{0.3}I_{0.7}$)$_3$ (Pm-3m space group) collected at 295, 318, 333, and 353 K obtained from refinements of synchrotron XRD pattern using the Rietveld method. **d)** Compressibility *K* of the lattice parameters in the cubic phase. Given the cubic structure, the compressibility is equal along all three axes.

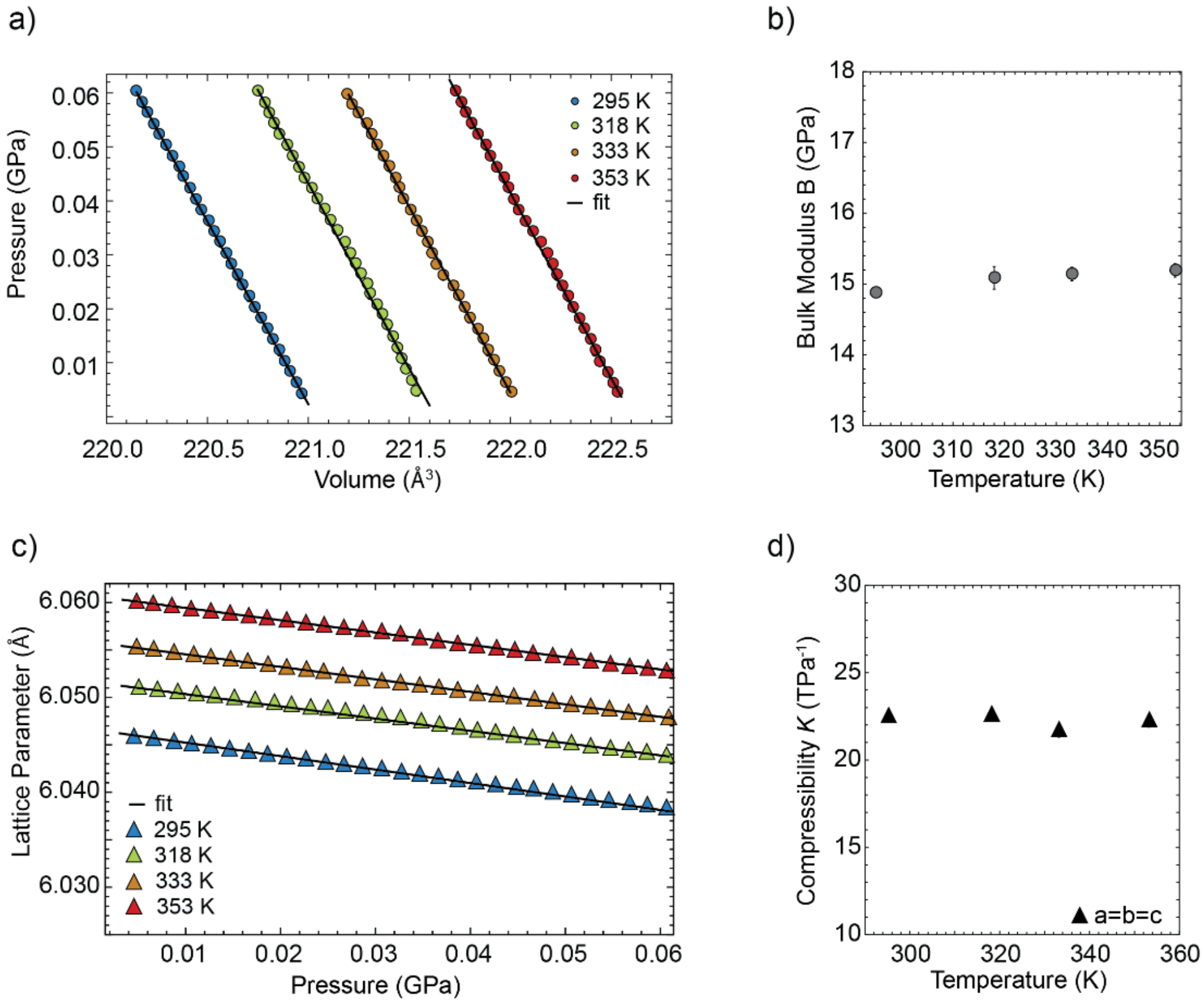

**Figure S5**. Unit-cell volume changes with pressure for $MAPb(Br_{0.7}I_{0.3})_3$ in the cubic Pm-3m phase and the corresponding fit to equation 1 of the main text. **b)** Bulk moduli obtained from the data shown in (a) and reported as a function of temperature in the cubic phase show no temperature dependence. **c)** Pressure-dependent lattice parameters of cubic $MAPb(Br_{0.7}I_{0.3})_3$ (Pm-3m space group) collected at 295, 318, 333, and 353 K obtained from refinements of synchrotron XRD pattern using the Rietveld method. **d)** Compressibility *K* of the lattice parameters in the cubic phase. Given the cubic structure, the compressibility is equal along all three axes.

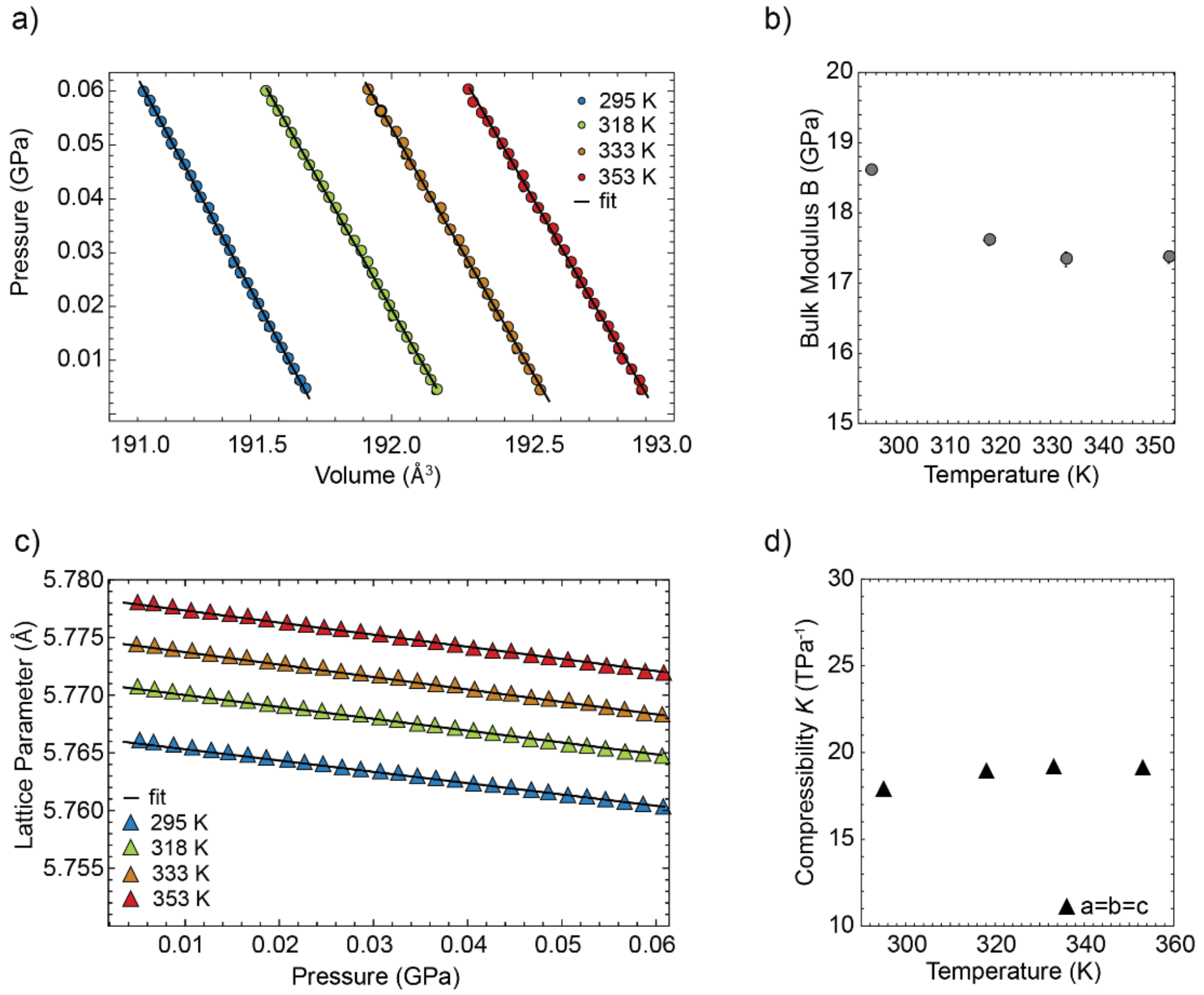

**Figure S6**. Unit-cell volume changes with pressure for $MAPb(Br_{0.3}Cl_{0.7})_3$ in the cubic Pm-3m phase and the corresponding fit to equation 1 of the main text. **b)** Bulk moduli obtained from the data shown in (a) and reported as a function of temperature in the cubic phase show no temperature dependence. **c)** Pressure-dependent lattice parameters of cubic $MAPb(Br_{0.3}Cl_{0.7})_3$ (Pm-3m space group) collected at 295, 318, 333, and 353 K obtained from refinements of synchrotron XRD pattern using the Rietveld method. **d)** Compressibility *K* of the lattice parameters in the cubic phase. Given the cubic structure, the compressibility is equal along all three axes.

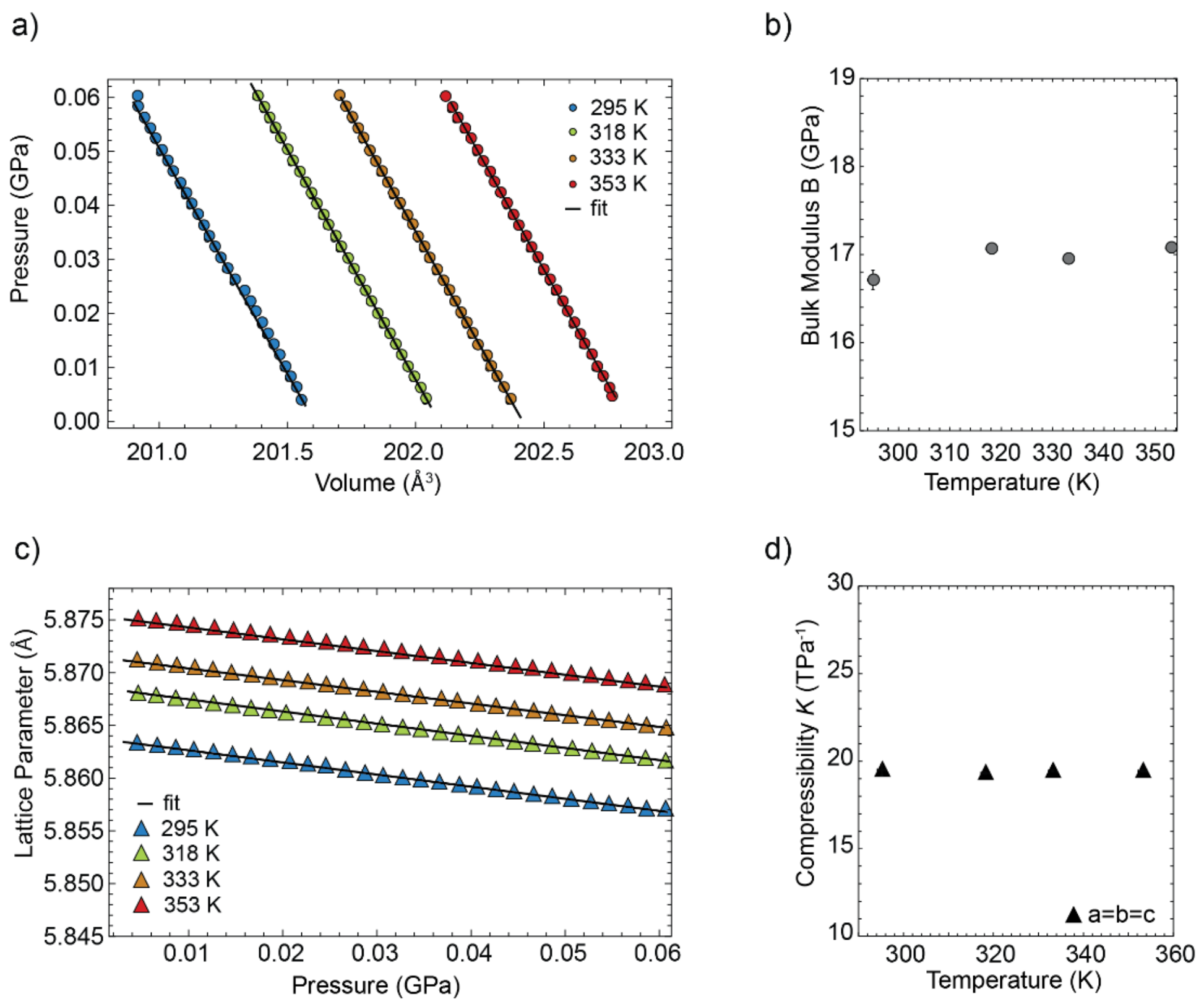

**Figure S7**. Unit-cell volume changes with pressure for $MAPb(Br_{0.7}Cl_{0.3})_3$ in the cubic Pm-3m phase and the corresponding fit to equation 1 of the main text. **b)** Bulk moduli obtained from the data shown in (a) and reported as a function of temperature in the cubic phase show no temperature dependence. **c)** Pressure-dependent lattice parameters of cubic $MAPb(Br_{0.7}Cl_{0.3})_3$ (Pm-3m space group) collected at 295, 318, 333, and 353 K obtained from refinements of synchrotron XRD pattern using the Rietveld method. **d)** Compressibility *K* of the lattice parameters in the cubic phase. Given the cubic structure, the compressibility is equal along all three axes.

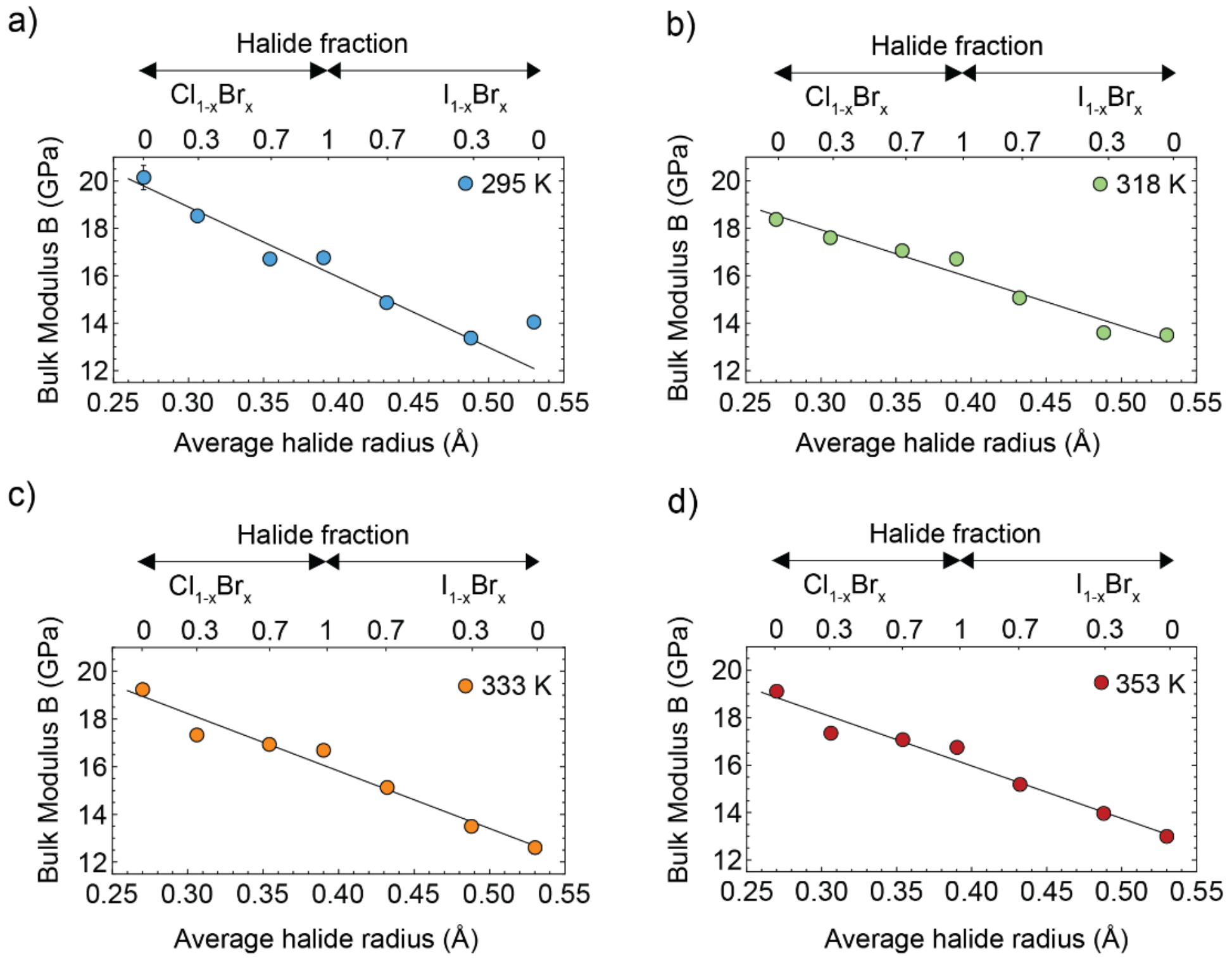

**Figure S8**. Linear fit of the bulk modulus as a function of the average halide radius at **a)** 295 K, **b)** 318 K, **c)** 333 K and **d)** 353K for single halide and mixed-halide compositions. Bulk moduli of compositions with intermediate average halide radius can be calculated using $B = 25.9 - 24.1x$ at room temperature, $B = 24 - 20.2x$ at 318K, $B = 25.4 - 24.1x$ at 333K, $B = 24.8 - 22.1x$ at 353K, where B is the bulk modulus and $x$ the average halide radius.

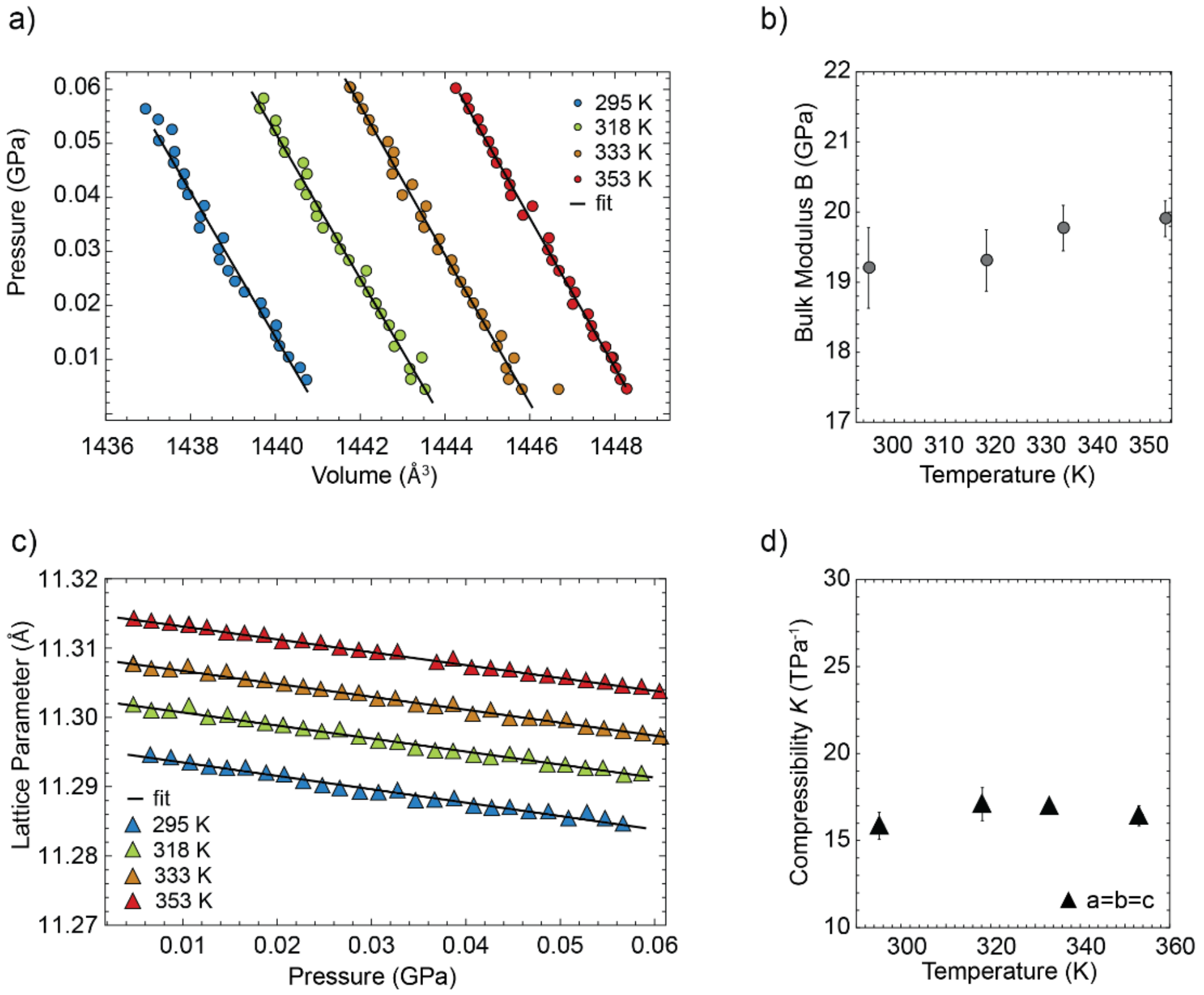

**Figure S9**. Unit-cell volume changes with pressure for $Cs_2AgBiBr_6$ in the cubic Fm-3m phase and the corresponding fit to equation 1 of the main text. **b)** Bulk moduli obtained from the data shown in (a) and reported as a function of temperature in the cubic phase show no temperature dependence. **c)** Pressure-dependent lattice parameters of cubic $Cs_2AgBiBr_6$ (Fm-3m space group) collected at 295, 318, 333, and 353 K obtained from refinements of synchrotron XRD pattern using the Rietveld method. **d)** Compressibility *K* of the lattice parameters in the cubic phase. Given the cubic structure, the compressibility is equal along all three axes.

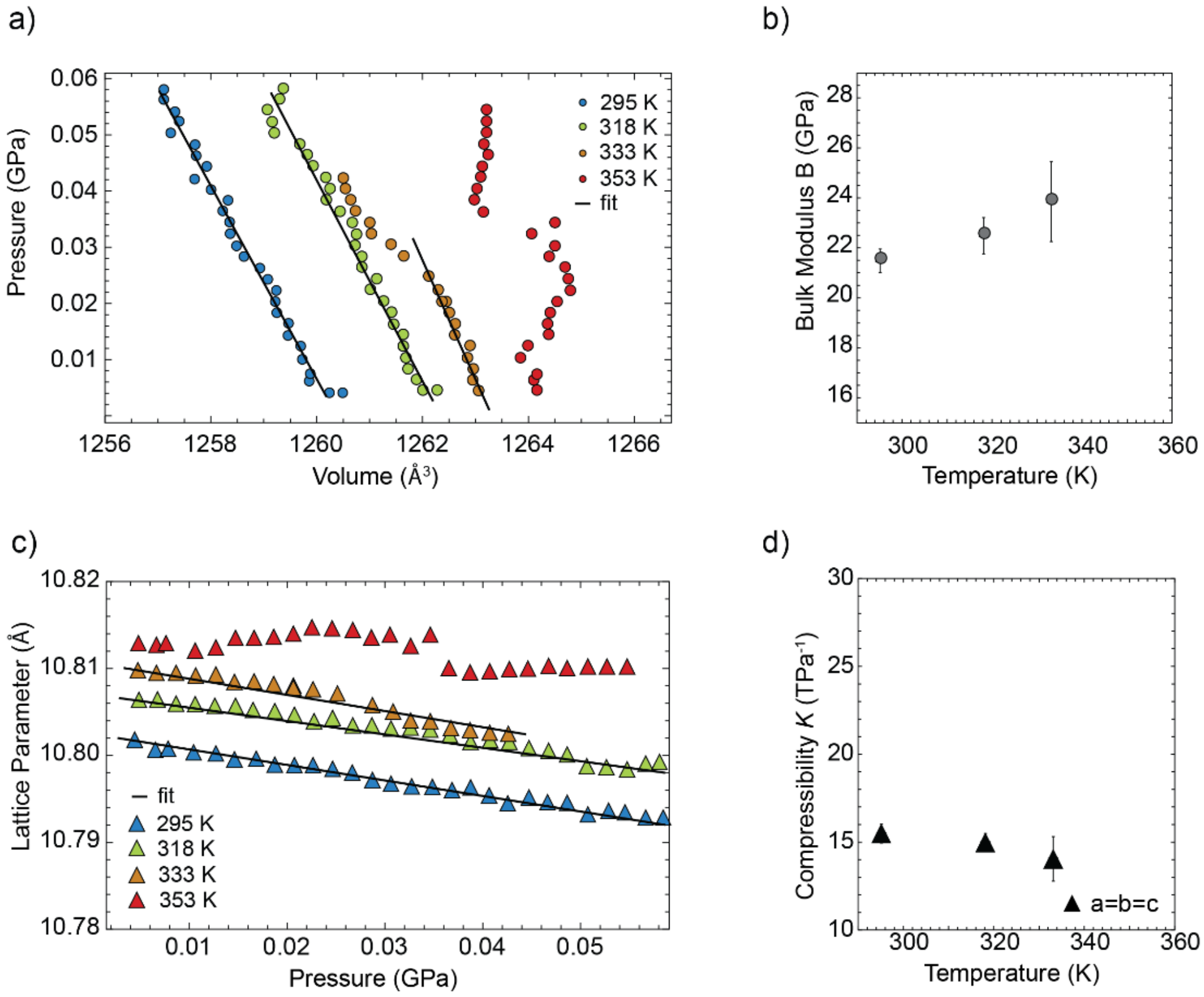

**Figure S10**. Unit-cell volume changes with pressure for $Cs_2AgBiCl_6$ in the cubic Fm-3m phase and the corresponding fit to equation 1 of the main text. **b)** Bulk moduli obtained from the data shown in (a) and reported as a function of temperature in the cubic phase show no temperature dependence. Due to the deviation in the linearity of P-V curve, we only fit the first half of the curve at 333 K, whereas we are not able to fit the P-V curve at 353 K. **c)** Pressure-dependent lattice parameters of cubic $Cs_2AgBiCl_6$ (Fm-3m space group) collected at 295, 318, 333, and 353 K obtained from refinements of synchrotron XRD pattern using the Rietveld method. **d)** Compressibility $K$ of the lattice parameters in the cubic phase. Given the cubic structure, the compressibility is equal along all the three axes. Due to the deviation in the linearity of P-V curve at 353 K we do not report the corresponding compressibility.

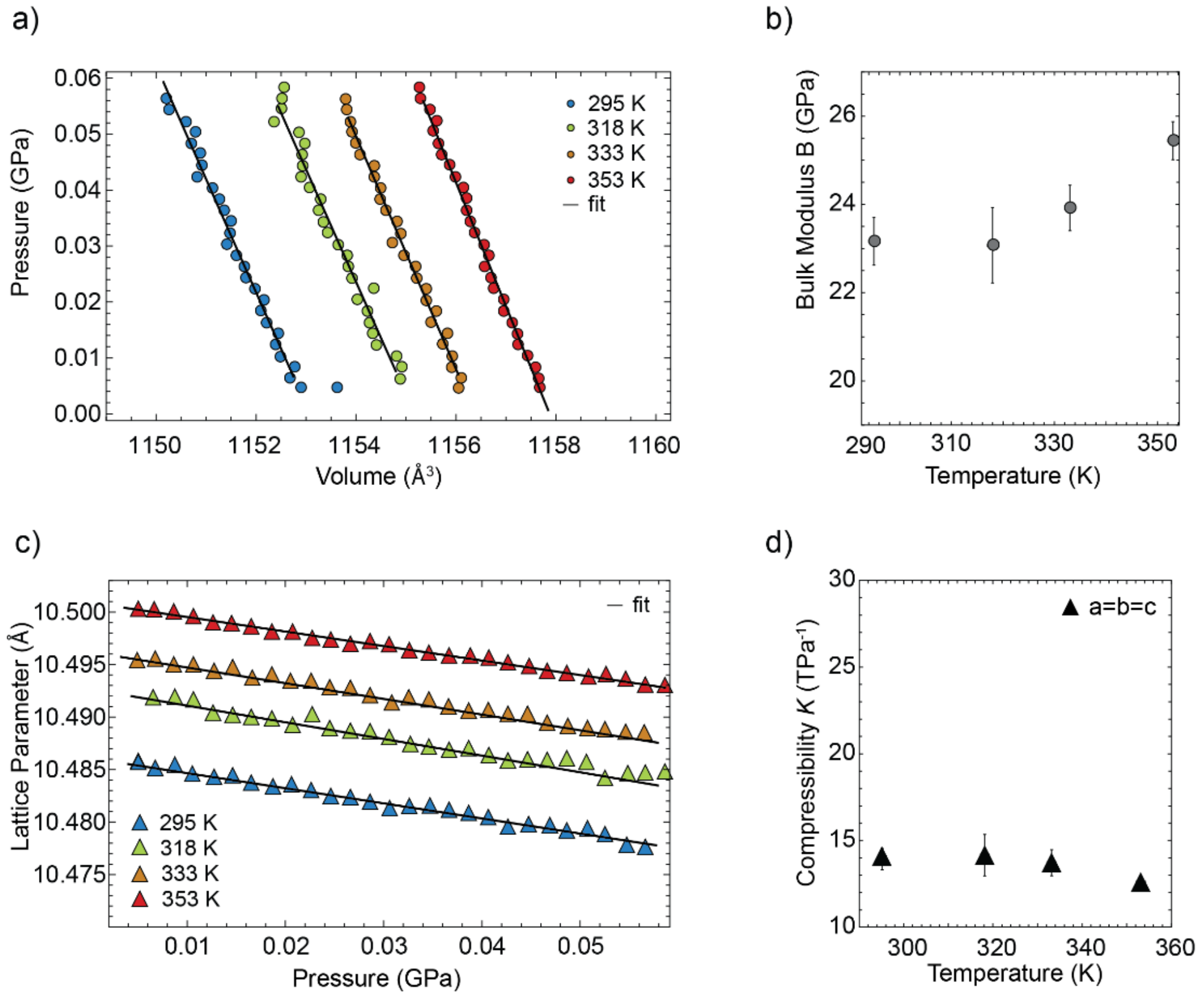

**Figure S11**. Unit-cell volume changes with pressure for $Cs_2AgInCl_6$ in the cubic Fm-3m phase and the corresponding fit to equation 1 of the main text. **b)** Bulk moduli obtained from the data shown in (a) and reported as a function of temperature in the cubic phase show no temperature dependence. **c)** Pressure-dependent lattice parameters of cubic $Cs_2AgInCl_6$ (Fm-3m space group) collected at 295, 318, 333, and 353 K obtained from refinements of synchrotron XRD pattern using the Rietveld method. **d)** Compressibility *K* of the lattice parameters in the cubic phase. Given the cubic structure, the compressibility is equal along all three axes.

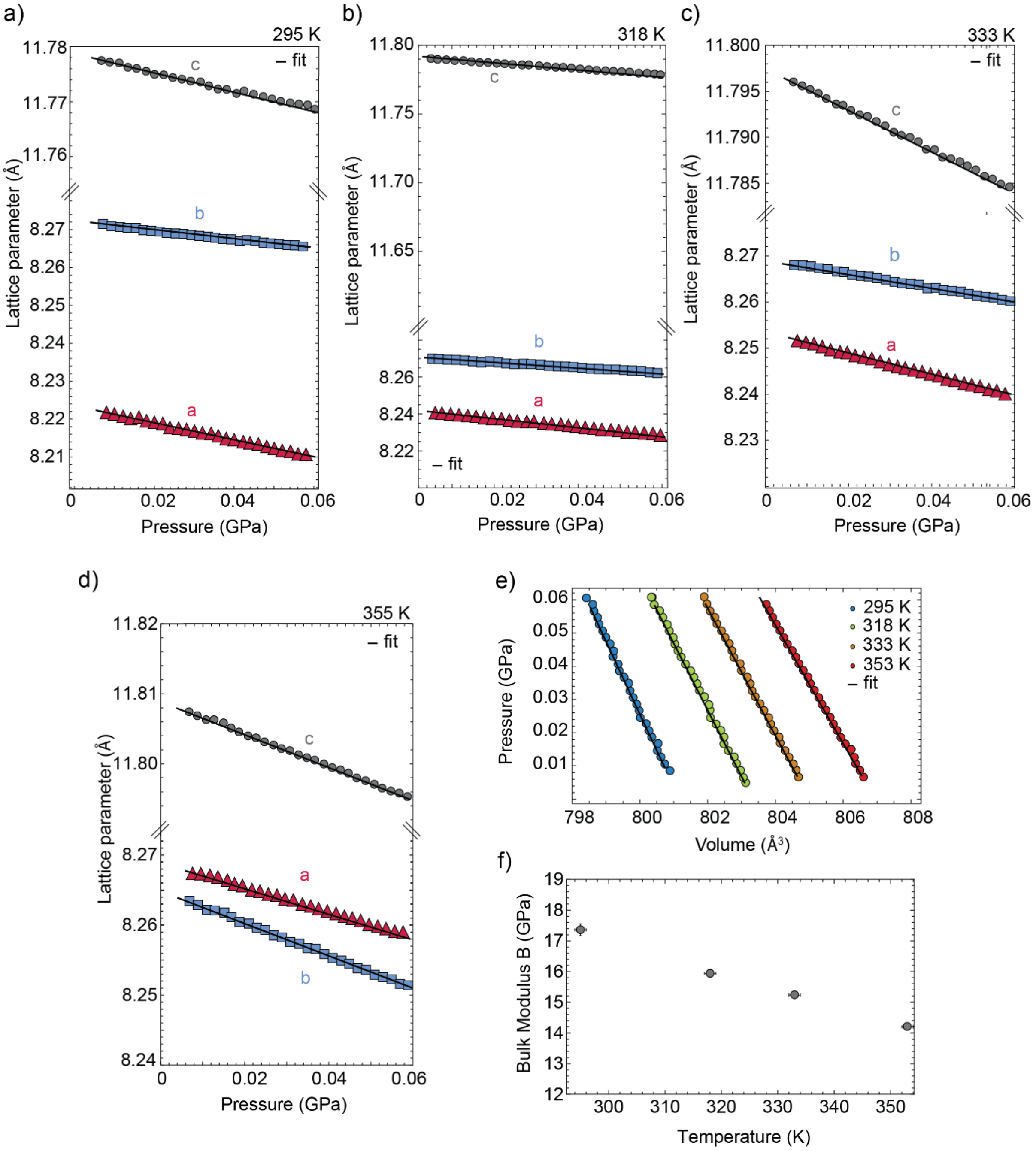

**Figure S12**. Pressure-dependent lattice parameters of the orthorhombic $CsPbBr_3$ (Pbnm space group) obtained from refinements of synchrotron XRD pattern at **a)** 295 K, **b)** 318 K, **c)** 333 K and **d)** 355 K using the Rietveld method. At 355 K, the lattice parameter *a* swap with that of *b*. **e)** Unit-cell volume changes with pressure for $CsPbBr_3$ in the orthorhombic Pbnm phase. **f)** Bulk modulus of $CsPbBr_3$ as a function of temperature obtained from the fit of the volume changes with pressure to equation 1 of the main text. Compressibility *K* is shown in **Figure 3b** in the main text.

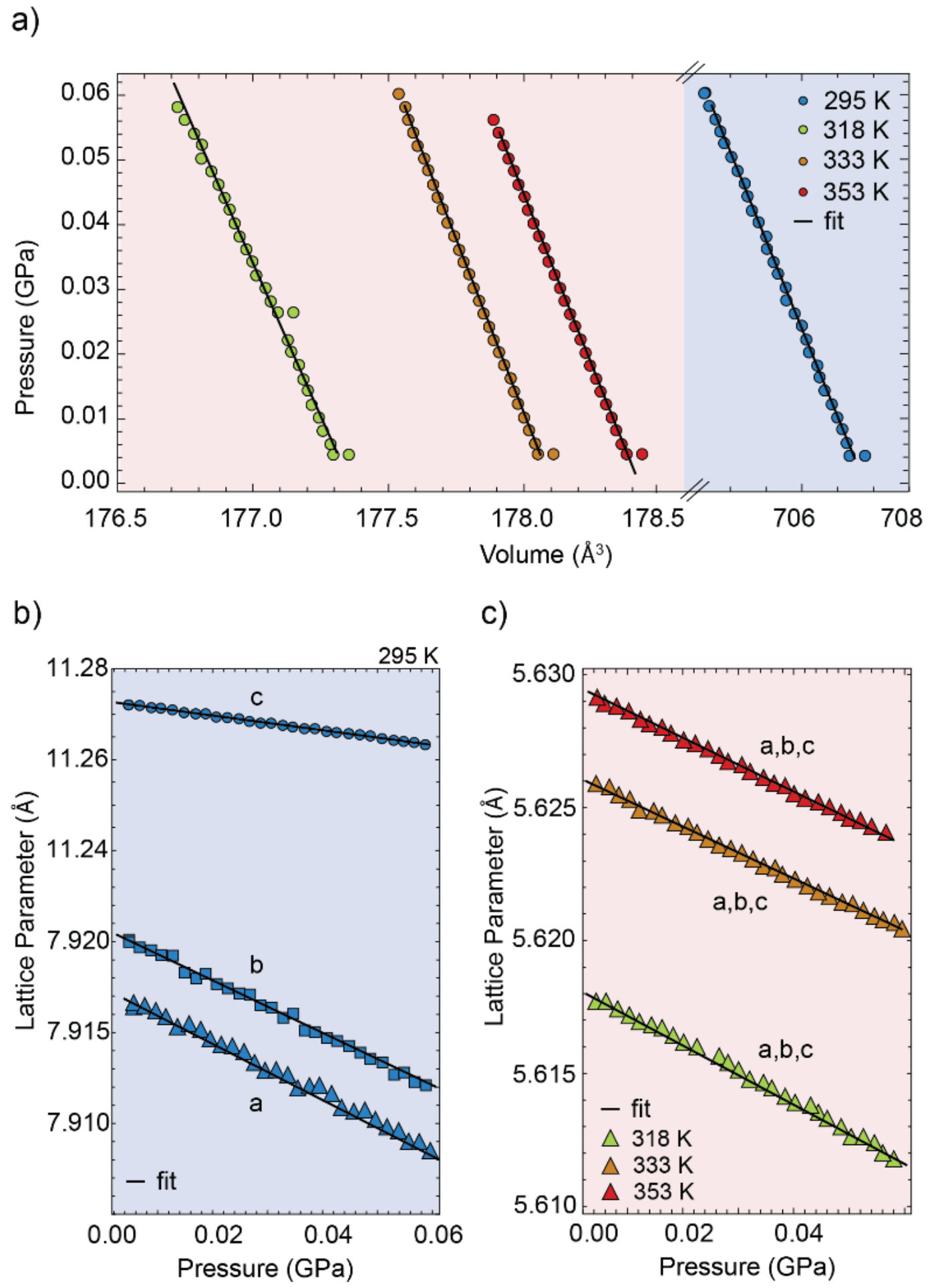

**Figure S13. a)** Unit-cell volume changes with pressure for $CsPbCl_3$ in the cubic Pm-3m phase (highlighted in red) and in the orthorhombic Pbnm phase (highlighted in blue) and the corresponding fit to equation 1 of the main text. **b)** Pressure-dependent lattice parameters of the orthorhombic $CsPbCl_3$ (Pbnm space group) collected at 295 K and **c)** of cubic Pm-3m phase at 318, 333, and 353 K obtained from refinements of synchrotron XRD pattern using the Rietveld method. Compressibility $K$ is shown in **Figure 3b** in the main text.

**Supporting Note 1. Fit for determining the bulk modulus as a function of temperature**

For all the elpasolite, single halide (iodide, bromide, chloride) and mixed-halide perovskite series $MAPb(Cl_{1-x}Br_x)_3$ and $MAPb(I_{1-x}Br_x)_3$, we estimate the bulk moduli (B) by fitting the following function to the pressure-volume trends (**Figure 1c–d** and **Figure S1–S7**) at 295, 318, 333 and 353K.

$$B = -\frac{\Delta P}{\Delta V} V \tag{2}$$

with $\Delta P/\Delta V$ the partial derivative of pressure with respect to volume, and $V$ the volume at ambient pressure. The results are reported below in **Table S1**.

| | $B$ (GPa) at 295 K | $B$ (GPa) at 318 K | $B$ (GPa) at 333 K | $B$ (GPa) at 353 K |
|---|---|---|---|---|
| **$MAPbI_3$** | 14.22 | 13.67 | 12.77 | 13.15 |
| **$MAPbBr_3$** | 16.94 | 16.87 | 16.85 | 16.91 |
| **$MAPbCl_3$** | 20.09 | 18.78 | 19.07 | 18.58 |
| **$MAPb(Br_{0.3}I_{0.7})_3$** | 13.53 | 13.78 | 13.66 | 14.13 |
| **$MAPb(Br_{0.7}I_{0.3})_3$** | 15.03 | 15.25 | 15.30 | 15.36 |
| **$MAPb(Br_{0.3}Cl_{0.7})_3$** | 18.77 | 17.77 | 17.50 | 17.53 |
| **$MAPb(Br_{0.7}Cl_{0.3})_3$** | 16.87 | 17.22 | 17.11 | 17.24 |
| **$CsPbBr_3$** | 17.53 | 16.10 | 15.40 | 14.36 |
| **$CsPbCl_3$** | 19.07 | 16.82 | 19.16 | 18.79 |
| **$Cs_2AgBiBr_6$** | 19.35 | 19.46 | 19.93 | 20.06 |
| **$Cs_2AgBiCl_6$** | 21.63 | 22.65 | 15.92 | - |
| **$Cs_2AgInCl_6$** | 23.31 | 23.23 | 24.07 | 25.59 |

**Table S1.** Bulk moduli, B, as a function of temperature for all compositions at 295 K, 318 K, 333 K, and 353 K.

**Supporting Note 2. Axes-dependent and volumetric compressibility as a function of temperature**

Considering that the elastic properties of non-cubic perovskites, such as tetragonal $MAPbI_3$, are in general anisotropic and therefore depend on crystallographic direction, we derived the compressibility of the specific crystal axes by monitoring their compression as a function of applied pressure (*e.g.*, for the compressibility along a-axis):

$$K_a = -\frac{(a_f - a_0)}{a_f}P \tag{3}$$

with $a$ the lattice parameter, and $a_f - a_0$ the change in lattice parameter between the lattice parameter at the final ($a_f$) and initial ($a_0$) pressure. Similarly, the compressibility for $b$- and $c$-axis, for non-cubic perovskites, can be determined. The results are reported in **Table S2-S4**. The volume compressibility, $K_V$, i.e., the susceptibility of a material to compress upon external applied pressure, is inversely proportional to $B$. $K_V$ is given by

$$K_V = K_a + K_b + K_c \tag{4}$$

i.e., $K_V$ is the sum of the compressibility along each axis. Results are reported in **Table S5**.

| | **$K_x$ (TPa$^{-1}$) at 295 K** | | **$K_x$ (TPa$^{-1}$) at 318 K** | | **$K_x$ (TPa$^{-1}$) at 333 K** | | **$K_x$ (TPa$^{-1}$) at 353 K** | |
|---|---|---|---|---|---|---|---|---|
| **$CsPbBr_3$** | $a$ | 27.75 | $a$ | 27.65 | $a$ | 27.91 | $a$ | 21.52 |
| | $b$ | 13.89 | $b$ | 17.01 | $b$ | 18.71 | $b$ | 28.61 |
| | $c$ | 15.77 | $c$ | 17.81 | $c$ | 18.68 | $c$ | 19.92 |
| **$CsPbCl_3$** | $a$ | 19.21 | | | | | | |
| | $b$ | 19.12 | - | | - | | - | |
| | $c$ | 14.49 | | | | | | |

**Table S2.** Compressibility along the a, b, and c-axis, $K_x$ $(with\, x = a, b, c)$, for the orthorhombic compositions at 295 K, 318 K, 333 K, and 353 K.

| | $K_x$ (TPa$^{-1}$) at 295 K | | $K_x$ (TPa$^{-1}$) at 318 K | | $K_x$ (TPa$^{-1}$) at 333 K | $K_x$ (TPa$^{-1}$) at 353 K |
|---|---|---|---|---|---|---|
| $MAPbI_3$ | $a = b$ | 28.38 | $a = b$ | 28.98 | - | - |
| | $c$ | 13.93 | $c$ | 15.64 | | |

**Table S3.** Compressibility along the a, b, and c-axis, $K_x$ $(with\, x = ab, c)$, for the tetragonal compositions at 295 K, 318 K, 333 K, and 353 K.

| | $K_a$ (TPa$^{-1}$) at 295 K | $K_a$ (TPa$^{-1}$) at 318 K | $K_a$ (TPa$^{-1}$) at 333 K | $K_a$ (TPa$^{-1}$) at 353 K |
|---|---|---|---|---|
| $MAPbI_3$ | - | - | 26.24 | 25.48 |
| $MAPbBr_3$ | 19.73 | 19.84 | 19.83 | 19.80 |
| $MAPbCl_3$ | 12.07 | 17.82 | 17.55 | 18.01 |
| $MAPb(Br_{0.3}I_{0.7})_3$ | 24.75 | 24.26 | 24.54 | 23.70 |
| $MAPb(Br_{0.7}I_{0.3})_3$ | 22.29 | 21.96 | 21.89 | 21.81 |
| $MAPb(Br_{0.3}Cl_{0.7})_3$ | 17.84 | 18.83 | 19.13 | 19.09 |
| $MAPb(Br_{0.7}Cl_{0.3})_3$ | 19.84 | 19.44 | 19.56 | 19.39 |
| $CsPbCl_3$ | - | 17.76 | 17.49 | 17.95 |
| $Cs_2AgBiBr_6$ | 17.29 | 17.19 | 16.79 | 16.68 |
| $Cs_2AgBiCl_6$ | 15.46 | 14.77 | 21.02 | 16.00 |
| $Cs_2AgInCl_6$ | 14.34 | 14.40 | 13.89 | 13.06 |

**Table S4.** Compressibility along the a-axis, $K_a$, for the cubic compositions at 295 K, 318 K, 333 K, and 353 K.

| | $K_V$ (TPa$^{-1}$) | $K_V$ (TPa$^{-1}$) | $K_V$ (TPa$^{-1}$) | $K_V$ (TPa$^{-1}$) |
|---|---|---|---|---|

| | at 295 K | at 318 K | at 333 K | at 353 K |
|---|---|---|---|---|
| **$MAPbI_3$** | 70.69 | 73.60 | 78.72 | 76.44 |
| **$MAPbBr_3$** | 59.19 | 59.52 | 59.49 | 59.40 |
| **$MAPbCl_3$** | 36.21 | 53.46 | 52.65 | 54.03 |
| **$MAPb(Br_{0.3}I_{0.7})_3$** | 74.25 | 72.78 | 73.62 | 71.10 |
| **$MAPb(Br_{0.7}I_{0.3})_3$** | 66.87 | 65.88 | 65.67 | 65.43 |
| **$MAPb(Br_{0.3}Cl_{0.7})_3$** | 53.52 | 56.49 | 57.39 | 57.27 |
| **$MAPb(Br_{0.7}Cl_{0.3})_3$** | 59.52 | 58.32 | 19.56 | 58.68 |
| **$CsPbBr_3$** | 57.5 | 62.47 | 65.3 | 70.05 |
| **$CsPbCl_3$** | 52.82 | 53.29 | 52.49 | 53.86 |
| **$Cs_2AgBiBr_6$** | 50.64 | 50.75 | 49.84 | 49.68 |
| **$Cs_2AgBiCl_6$** | 45.69 | 43.00 | 60.18 | - |
| **$Cs_2AgInCl_6$** | 42.38 | 41.68 | 41.11 | 38.81 |

**Table S5.** Volumetric compressibility $K_V$ for the compositions studied at 295 K, 318 K, 333 K, and 353 K.

**Supporting Note 3. Calculation of the thermal expansivity**

The thermal expansion coefficient, $\alpha_L$, of a material is defined as the relative expansion in one dimension per unit of temperature:

$$\alpha_L = \frac{1}{L}\frac{\Delta L}{\Delta T} \tag{5}$$

For solids $\alpha_L$ is typically temperature-independent. So, we expect a linear dependence of the lattice parameters, $L = a$, $b$, and $c$, over our range of measured temperatures. As such for $L = a$:

$$\alpha_\mathrm{a} = \frac{1}{a}\frac{\Delta a}{\Delta T} \tag{6}$$

Rewriting eq. (6) provides:

$$\frac{\Delta a}{a} = \alpha_\mathrm{a}\Delta T \tag{7}$$

For an $\alpha_\mathrm{a}$ that is independent to temperature this yields:

$$a(T) = a_0 + \alpha_\mathrm{a} a_0 T \tag{8}$$

where $a_0$ is the lattice parameter extrapolated to T = 0 K. For some compositions we observe non-linear expansion along certain crystal axes. In this case a single value for $\alpha_\mathrm{a}$ does not suffice and integration of eq. (6) over the experimental temperature range is required:

$$a(T) = a_0 Exp\left[\int_{T_1}^{T_2} \alpha_\mathrm{a}(T)\, dT\right] \tag{9}$$

with a(T) a second order polynomial function ($a(T) = m_1T^2 + m_2T + a_0$) describing the lattice parameter as a function of temperature. Solving eq. (9) for $\alpha_\mathrm{a}$ gives:

$$\alpha_\mathrm{a}(T) = \frac{m_2 + 2m_1T}{a_0 + T(m_2 + m_1T)} \tag{10}$$

**Table S6** provides the optimized parameters for eq. (10).

| **i)** | **Orthorhombic** | | | |
|---|---|---|---|---|
| | Crystal axis | $m_1$ | $m_2$ | $a_0$ |
| **MAPbI$_3$** | $a$ (× $10^{-5}$ K$^{-1}$) | 3.81 | | |
| | $b$ (× $10^{-5}$ K$^{-1}$) | 2.02 | | |
| | $c$ (× $10^{-5}$ K$^{-1}$) | 5.81 | | |
| **MAPbBr$_3$** | $a$ (× $10^{-5}$ K$^{-1}$) | 9.04 | | |
| | $b$ (× $10^{-5}$ K$^{-1}$) | −2.79 | | |
| | $c$ (× $10^{-5}$ K$^{-1}$) | 1.00 | | |
| **MAPbCl$_3$** | a | $-8.45 \times 10^{-7}$ | $5.63 \times 10^{-4}$ | 11.15 |
| | $b$ | $1.86 \times 10^{-6}$ | $-3.05 \times 10^{-4}$ | 11.36 |
| | C | $8.60 \times 10^{-7}$ | $-3.26 \times 10^{-5}$ | 11.28 |
| **CsPbBr$_3$** | $a$ (× $10^{-5}$ K$^{-1}$) | 7.68 | | |
| | $b$ | $1.22 \times 10^{-6}$ | $-6.14 \times 10^{-4}$ | 8.34 |
| | $c$ (× $10^{-5}$ K$^{-1}$) | 3.51 | | |
| **CsPbCl$_3$** | $a$ | $-9.82 \times 10^{-7}$ | $9.18 \times 10^{-4}$ | 7.72 |
| | $b$ | $-4.61 \times 10^{-7}$ | $5.14 \times 10^{-4}$ | 11.14 |
| | $c$ | 1.41 | $-5.80 \times 10^{-4}$ | 7.96 |

| **ii)** | **Tetragonal** | | | |
|---|---|---|---|---|
| | Crystal axis | $m_1$ | $m_2$ | $a_0$ |
| **MAPbI$_3$** | $a = b$ | $1.03 \times 10^{-6}$ | $5.10 \times 10^{-5}$ | 8.76 |
| | $c$ (× $10^{-5}$ K$^{-1}$) | 0.34 | | |
| **MAPbBr$_3$** | $a = b$ (× $10^{-5}$ K$^{-1}$) | 6.99 | | |
| | $c$ | $-5.52 \times 10^{-6}$ | $1.63 \times 10^{-3}$ | 11.77 |

| **iii)** | **Cubic** (× $10^{-5}$ K$^{-1}$) |
|---|---|
| **$MAPbI_3$** | 4.04 |
| **$MAPbBr_3$** | 3.53 |
| **$MAPbCl_3$** | 3.51 |
| **$CsPbBr_3$** | 3.27** |
| **$CsPbCl_3$** | 3.02 |
| **$Cs_2AgBi(Br_{0.33}I_{0.67})_6$** | 5.08** |
| **$Cs_2AgBiBr_6$** | 2.86 |
| **$Cs_2AgBiCl_6$** | 2.42 |
| **$Cs_2AgInCl_6$** | 2.49 |
| **$Cs_2Ag(Bi_{0.5}In_{0.5})Br_6$** | 2.84** |
| **$Cs_2Ag(Bi_{0.5}Sb_{0.5})Br_6$** | 3.34** |
| **$Cs_2Ag(Bi_{0.9}Fe_{0.1})Br_6$** | 3.16** |

**Table S6.** Optimized parameters to describe α(T) for a given temperature range.

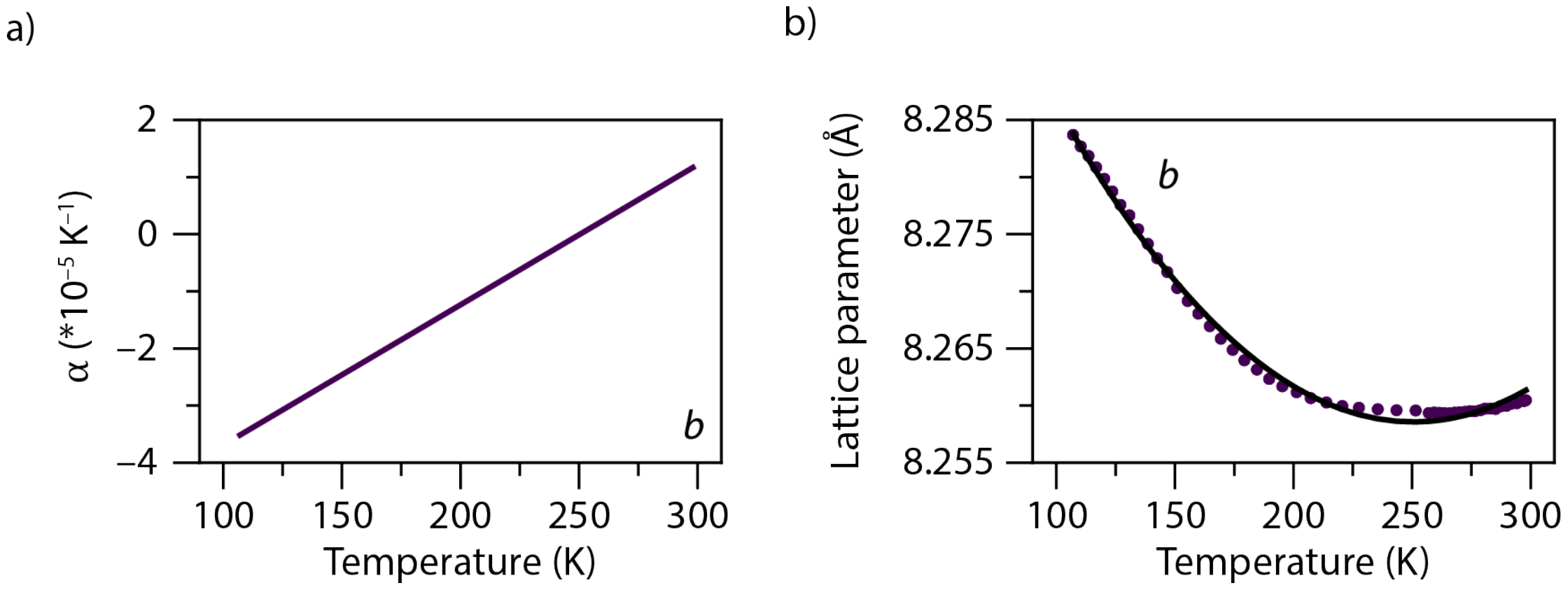

**Figure S16. a)** Non-linear $\alpha$ of the lattice parameter, $b$, of orthorhombic CsPbBr3 between 100 and 300 K. The optimized parameters for this plot are provided in **Table S6 i)**. **b)** Temperature dependent lattice parameter of the lattice parameter, $b$, of orthorhombic CsPbBr3 between 100 and 300 K (dots) and the polynomial fit (dotted line) to extract the temperature dependent $\alpha$ given in **a)**.

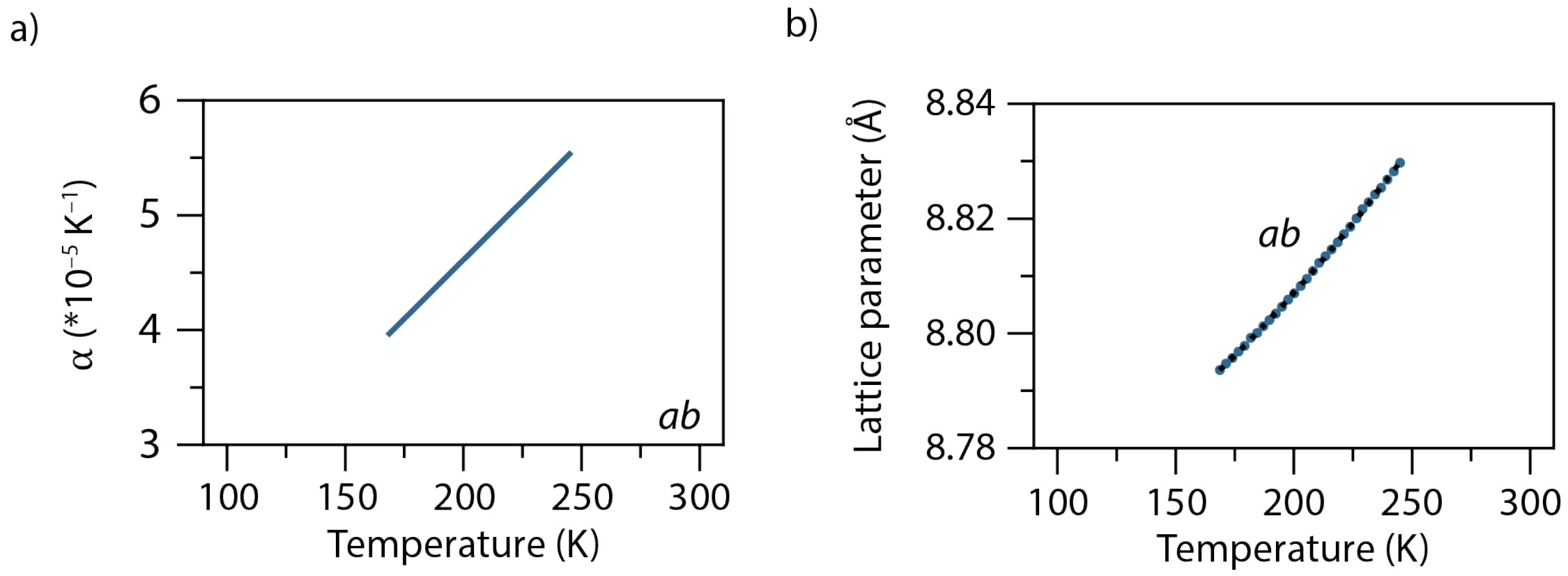

**Figure S17. a)** Non-linear $\alpha$ of the lattice parameter, *ab*, of tetragonal $MAPbI_3$ between 170 and 250 K. The optimized parameters for this plot are provided in **Table S6 ii)**. **b)** Temperature dependent lattice parameter of the lattice parameter, *ab*, of tetragonal $MAPbI_3$ between 170 and 250 K (dots) and the polynomial fit (dotted line) to extract the temperature dependent $\alpha$ given in **a)**.

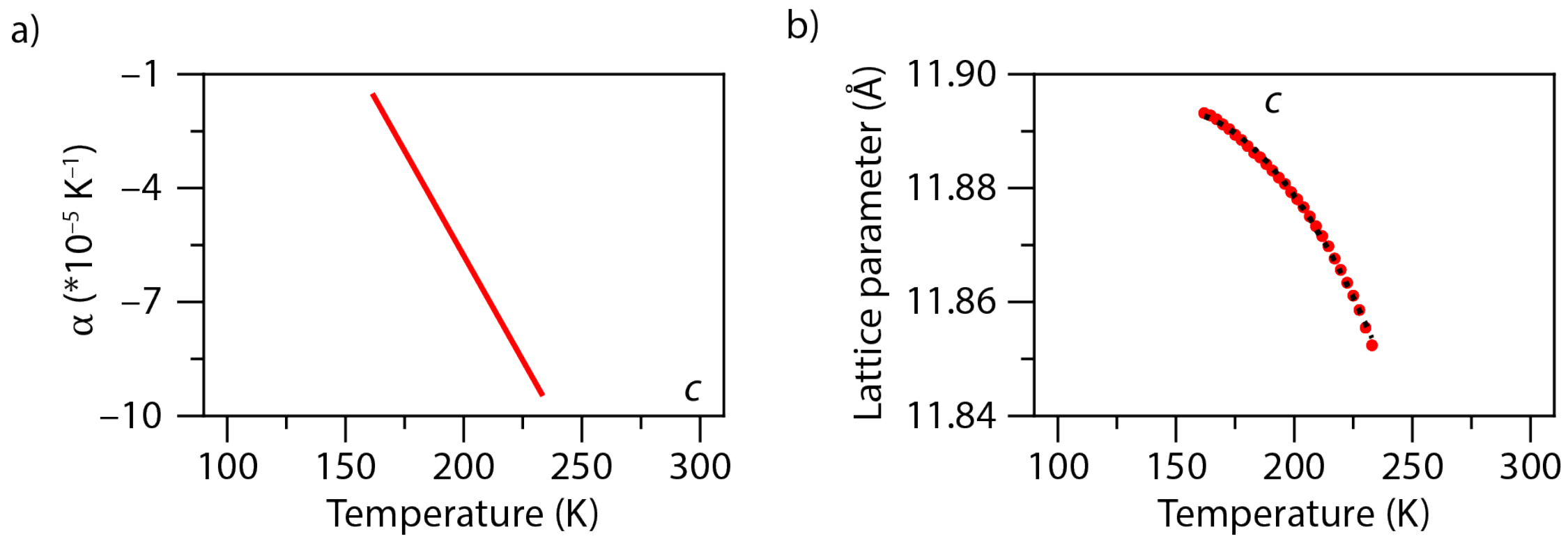

**Figure S18. a)** Non-linear $\alpha$ of the lattice parameter, *c*, of tetragonal $MAPbBr_3$ between 160 and 230 K. The optimized parameters for this plot are provided in **Table S6 ii)**. **b)** Temperature dependent lattice parameter of the lattice parameter, *c*, of tetragonal $MAPbBr_3$ between 160 and 230 K (dots) and the polynomial fit (dotted line) to extract the temperature dependent $\alpha$ given in **a)**.

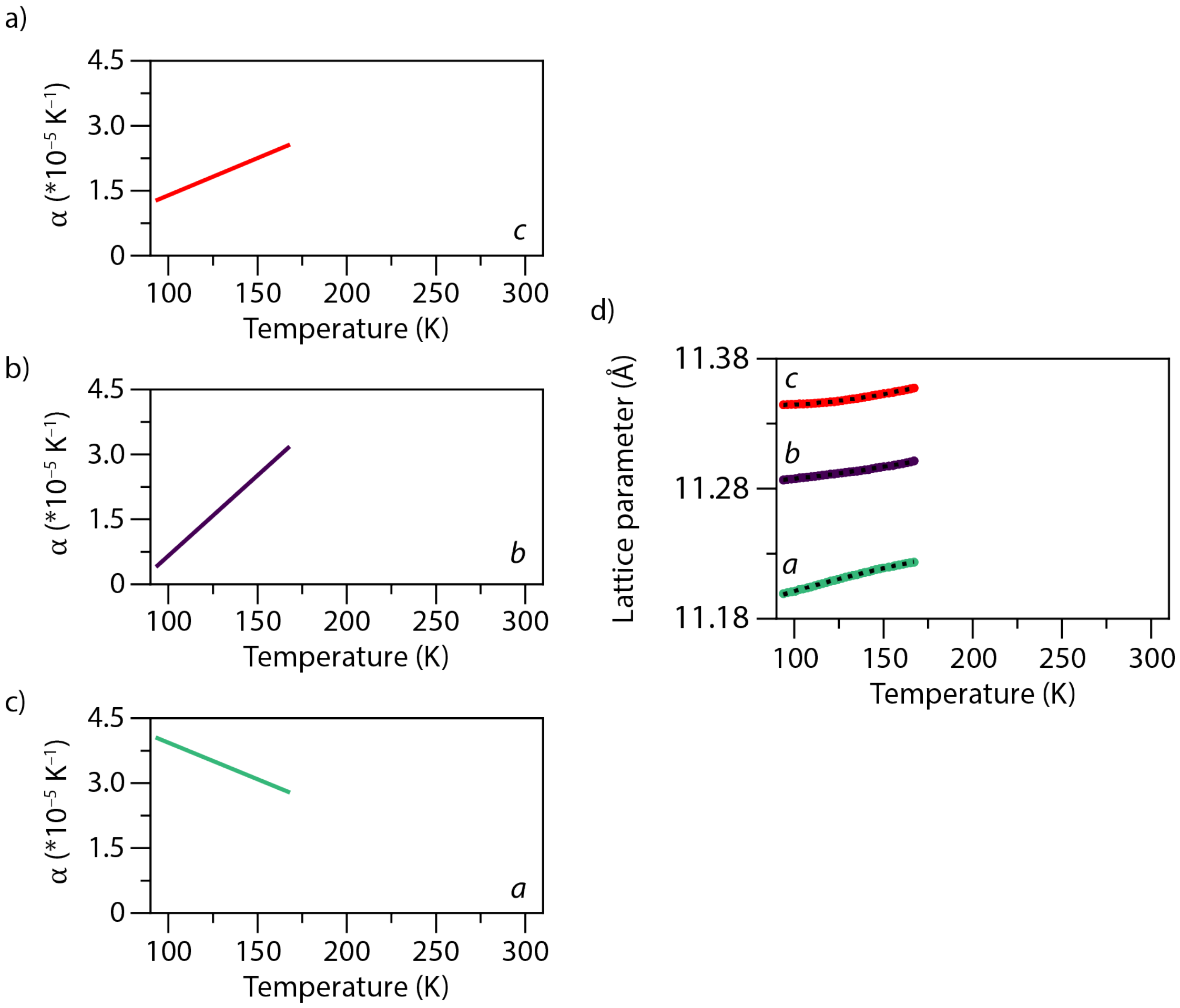

**Figure S19. a–c)** Non-linear $\alpha$ of the lattice parameters, *abc*, of orthorhombic $MAPbCl_3$ between 100 and 170 K. The optimized parameters for this plot are provided in **Table S6 i)**. **d)** Temperature dependent lattice parameter of the lattice parameter, *abc*, of orthorhombic $MAPbCl_3$ between 100 and 170 K (dots) and the polynomial fits (dotted line) to extract the temperature dependent $\alpha$ given in **a–c)**.

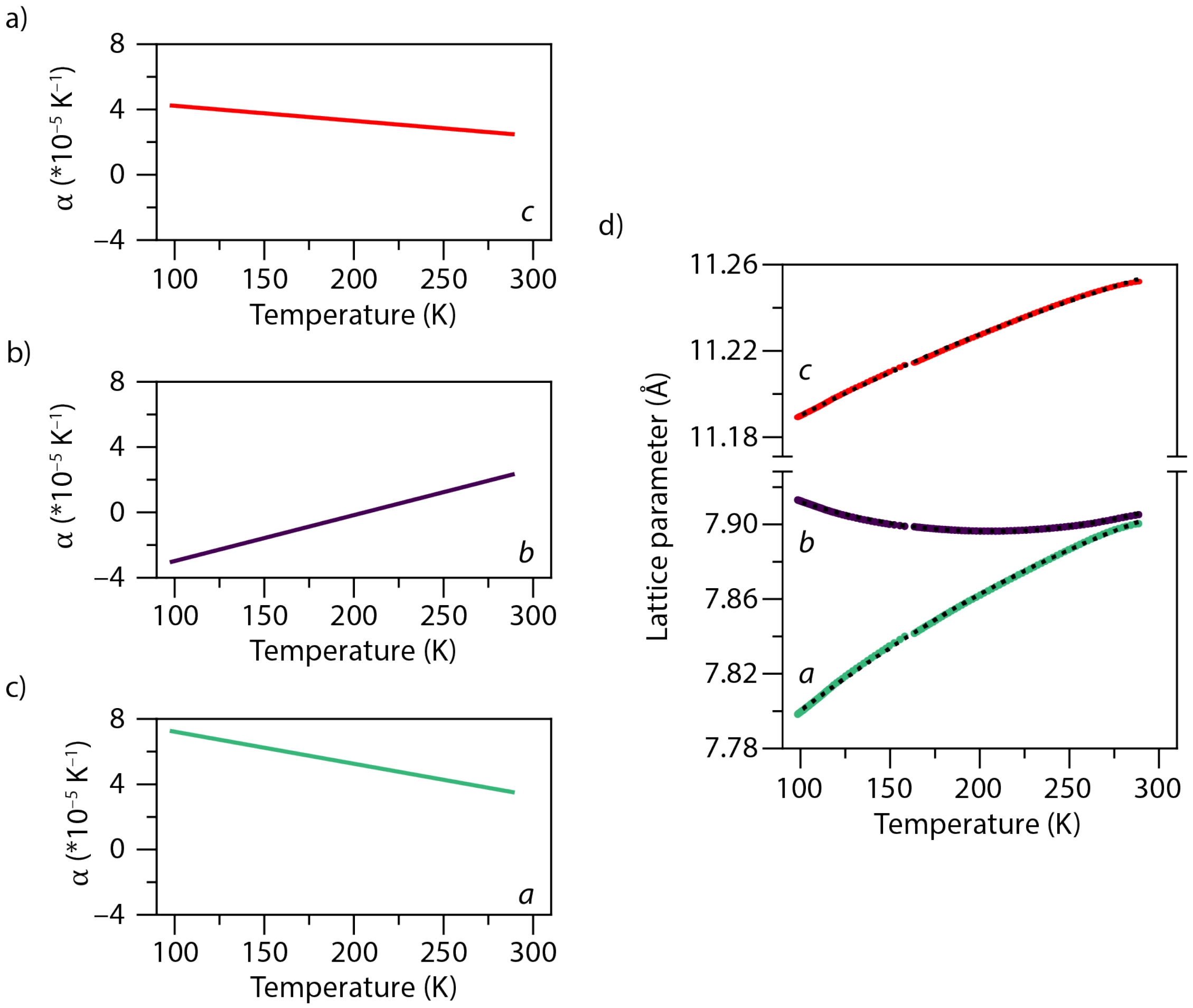

**Figure S20. a–c)** Non-linear $\alpha$ of the lattice parameters, *abc*, of orthorhombic $CsPbCl_3$ between 100 and 300 K. The optimized parameters for this plot are provided in **Table S6 i)**. **d)** Temperature dependent lattice parameter of the lattice parameter, *abc*, of orthorhombic $CsPbCl_3$ between 100 and 300 K (dots) and the polynomial fits (dotted line) to extract the temperature dependent $\alpha$ given in **a–c)**.

**Supporting Note 4. Lab-based temperature dependent X-ray diffraction**

First the temperature inside the cell was calibrated by measuring the lattice expansion of MgO over a temperature range of 300 K to 900 K. From literature the lattice parameter as a function of temperature is given by[5]:

$$a(T) = 4.2094 + 5.92 \times 10^{-5} \times T \quad (1)$$

with T the temperature in degrees Celsius. The lattice parameter of MgO as a function of temperature was determined using the same refinement method as described in the main text (**Figure S14a**). The experimental temperature was scaled with a factor of 1.05 to match the $\alpha$ of MgO reported in the literature (eq. (1)). No correction was performed to adjust for the change in sample position as a consequence of the expanding sample holder upon increasing the temperature.

To verify the reproducibility of our experiments, the $\alpha$ of $Cs_2AgBiBr_6$ (**Figure S14**, blue data points) and $CsPbBr_3$ (**Figure S15**) was determined and compared to the results from our experiments using synchrotron radiation. The discrepancy observed for these values can be attributed when considering the different experimental setups and conditions, such as the (mono-chromatic) light source, less accurately defined sample position, used detector, atmosphere, etc. We, however, note that quality of the lab-based data is still sufficient to apply the same refinement method as described in the main text. As such, we determined the $\alpha$ of $Cs_2Ag(Bi_{1-x}B_x)Br_6$ with ($B_x$ = $In_{0.5}$, $Sb_{0.5}$ and $Fe_{0.1}$) and $Cs_2AgBi(Br_{0.33}I_{0.67})_6$ (**Figure S14b** and **Table 1**).

a)

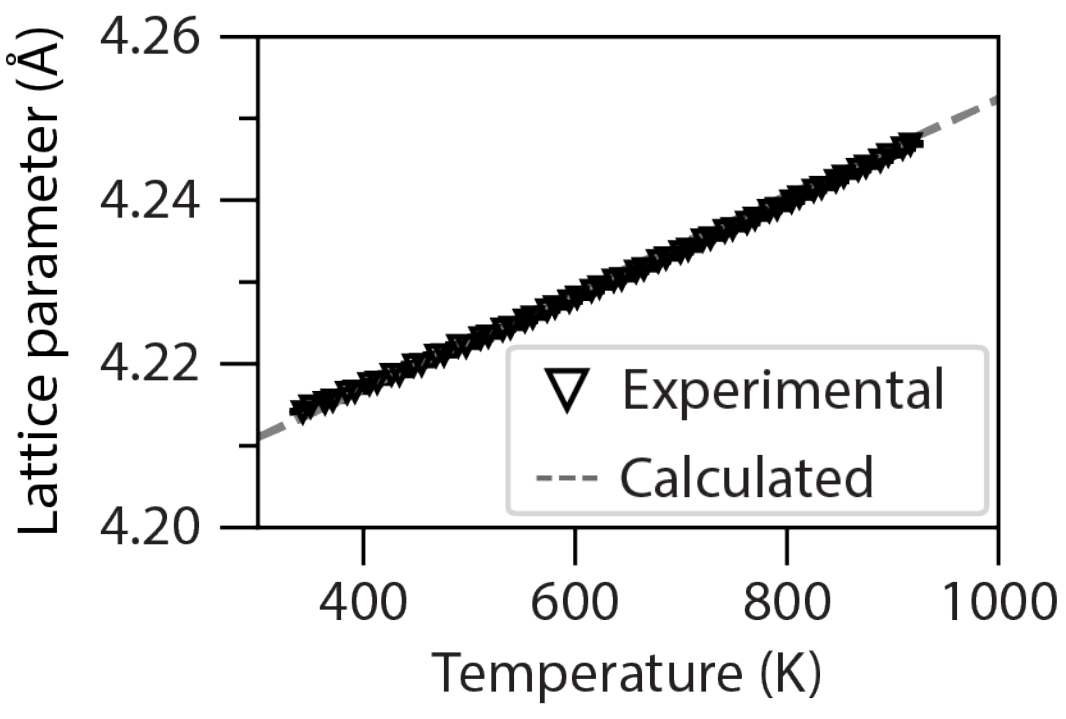

b)

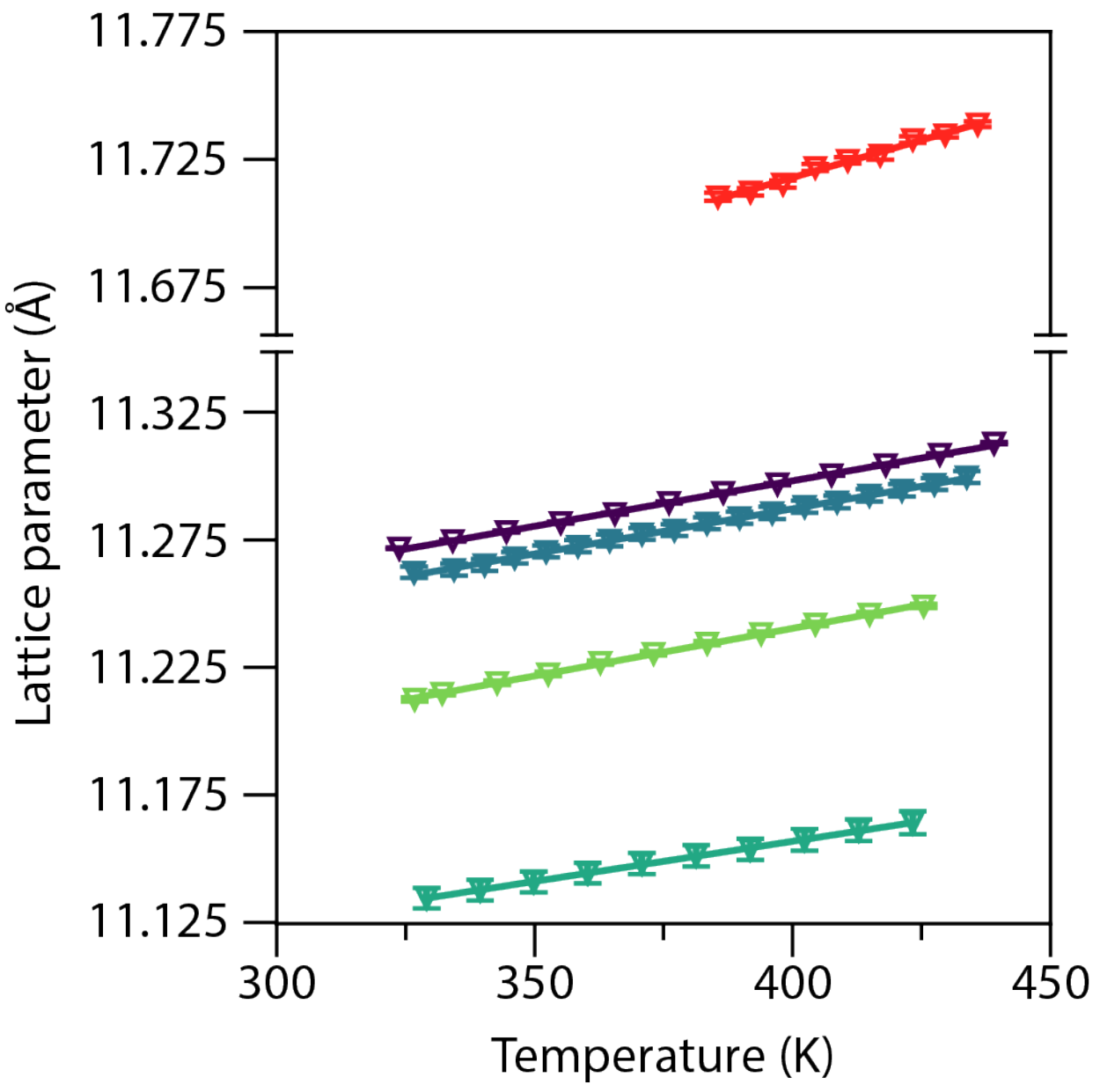

**Figure S14. a)** Temperature-dependent lattice parameter of MgO obtained using Rietveld refinement method of *in-situ* XRD patterns collected from 300 K to 900 K using a lab-based X-ray source. The dotted grey line represents eq. (1). **b)** Temperature-dependent lattice parameters of cubic alloyed elpasolites obtained using Rietveld refinement method of *in-situ* XRD patterns collected between 325 and 450 K using a lab-based X-ray source. From top to bottom the lines represent: $Cs_2AgBi(Br_{0.3}I_{0.7})_6$, $Cs_2AgBiBr_6$, $Cs_2Ag(Bi_{0.9}Fe_{0.1})Br_6$, $Cs_2Ag(Bi_{0.5}Sb_{0.5})Br_6$, and $Cs2A(Bi_{0.5}In_{0.5})Br_6$.

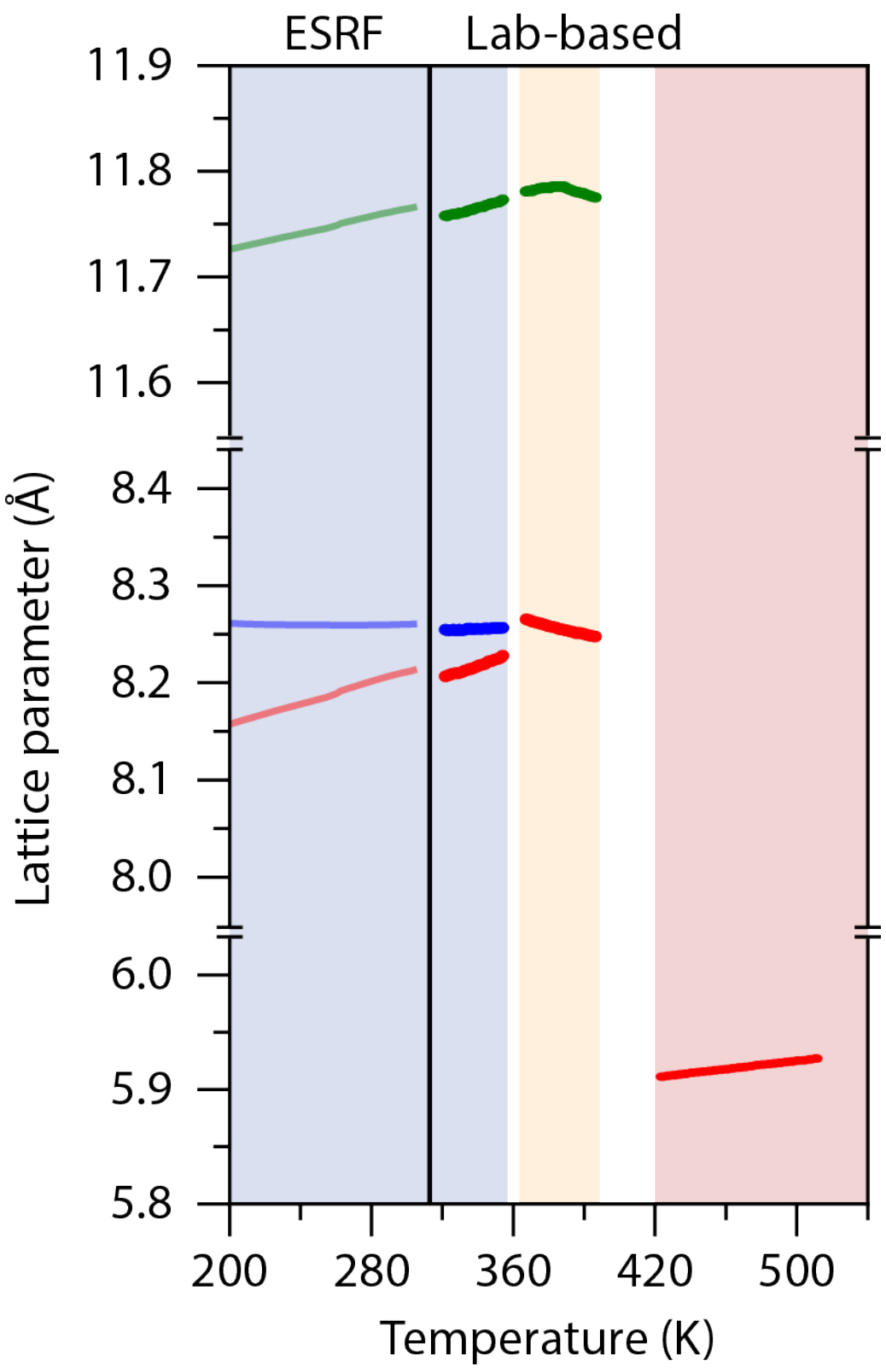

**Figure S15.** Lattice expansion of $CsPbBr_3$ over a temperature range of 200 to 540 K determined with synchrotron (<315 K) and lab-based X-ray sources (>315 K). For cubic CsPbBr3 (>420 K) the $\alpha$ was determined see **Table 1** in the main text.